\documentclass[twocolumn]{aastex62}

\newcommand{\del}{\mathbf{\nabla}}
\accepted{April 30, 2018}

\submitjournal{ApJ}

\shorttitle{Parker with Streaming}
\shortauthors{Heintz and Zweibel}

\usepackage{amsmath}
\usepackage{natbib}
\usepackage{xcolor}
\usepackage{bm}

\begin{document}

\title{The Parker Instability with Cosmic Ray Streaming}

\correspondingauthor{Evan Heintz}
\email{eheintz@wisc.edu, zweibel@astro.wisc.edu}

\author{Evan Heintz}
\affil{Department of Physics, University of Wisconsin - Madison, 475 North Charter Street, Madison, WI 53706, USA}

\author{Ellen G. Zweibel}
\affiliation{Department of Physics, University of Wisconsin - Madison, 475 North Charter Street, Madison, WI 53706, USA}
\affiliation{Department of Astronomy, University of Wisconsin - Madison, 475 North Charter Street, Madison, WI 53706, USA}

\begin{abstract}
Recent studies have found that cosmic ray transport plays an important role in feedback processes such as star formation and the launching of galactic winds. Although cosmic ray buoyancy is widely held to be a destabilizing force in galactic disks, the effect of cosmic ray transport on the stability of stratified systems has yet to be analyzed. We perform a stability analysis of
a stratified layer for three different cosmic ray transport models: decoupled (Classic Parker), coupled with $\gamma_c=4/3$ but not streaming (Modified Parker), and finally coupled with streaming at the Alfv\'{e}n speed. When the compressibility of the cosmic rays is decreased the system becomes much more stable, but the addition of cosmic ray streaming to the Parker Instability severely destabilizes it. Through comparison of these three cases and analysis of the work contributions for the perturbed quantities of each system, we demonstrate that cosmic ray heating of the gas is responsible for the destabilization of the system. We find that a 3D system is unstable over a larger range of wavelengths than the 2D system. Therefore, the Parker Instability with cosmic ray streaming may play an important role in cosmic ray feedback.
\end{abstract}


\section{Introduction} \label{sec:intro}
Cosmic rays are crucial for understanding many important phenomena in galaxies. Cosmic rays affect the dynamics, energy balance and chemistry of the interstellar medium \citep{Grenier2015,zweibelreview2017}. Cosmic rays are emerging as a major form of star formation
and possibly black hole feedback, along with radiation and thermal gas pressure. They may drive galactic outflows, help form large-scale structures like the Fermi Bubbles \citep{guofermi2012,yangfermi2012}, and play a role in galactic dynamos \citep[and references therein]{Hanasz2009}. In this paper, we are primarily concerned with the effects of cosmic rays on the launching of winds and the suppression of star formation, which we refer to collectively as feedback.

In order to better understand these phenomena, cosmic ray transport models have become a crucial area of study. There are many different treatments of cosmic ray transport, and it has been shown that feedback is sensitive to which one is adopted. For example, \cite{uhligwinds2012} and \cite{Salem2014} showed that if cosmic rays are locked to the thermal gas they are less effective in driving a wind than if they can diffuse \citep{boothwinds2013}, while \cite{ruszkowskiwinds2017} found that if the cosmic rays stream relative to the gas they are more effective at driving a wind and less effective in quenching star formation.

Feedback models with no cosmic ray transport show a
puffed-up galactic disc \citep{uhligwinds2012,ruszkowskiwinds2017} in which the star formation rate is reduced by lowering the mean thermal gas density through cosmic ray pressure support. It has been shown, however, that cosmic ray pressure supported systems can be unstable to Rayleigh-Taylor like modes known as the Parker Instability \citep{parkerinstability1966}.

Prior to its discovery in the galactic context,
the stability of plane stratified media in the presence of a horizontal magnetic field but no cosmic rays was analyzed by \cite{newcombinstability1961}. Newcomb showed that the stability criterion for this system reduces to the familiar Schwarzschild criterion:
\begin{equation}\label{eq:newcomb}
-\frac{g}{\rho}\frac{d\rho}{dz}-\frac{\rho g^2}{\gamma_g P_g} > 0
\end{equation}
for stability, where $g$ is the magnitude of the gravitational acceleration, $\rho$ is density, $P_g$ is gas pressure, and $\gamma_g$ is the adiabatic index. The magnetic field does not appear explicitly in eqn. (\ref{eq:newcomb}), but it affects stability through its effect on the stratification. Newcomb's criterion holds in the limit of infinitely long wavelength parallel to the magnetic field and infinitely
short wavelength in the horizontal direction perpendicular to it. 

In his original paper, \cite{parkerinstability1966} revisited the stability of this stratified system accounting for cosmic rays. Unstable perturbations are characterized by the magnetic field lines bending and the gas sliding down into the valleys. The system releases gravitational potential energy (which is destabilizing) but work must be done to compress the gas and resist magnetic tension (which are stabilizing). Parker assumed that the cosmic rays act as $\gamma=0$ fluid, which was consistent with the state of knowledge at that time, and affect the stratification only through their pressure gradient, which increases the scale height of the gas. In this case, cosmic rays are destabilizing.

Modal analysis shows that the growth time of the most unstable mode is of order the freefall time \citep{parkerinstability1966}, while nonlinear simulations show the growth of Rayleigh-Taylor like ``mushrooms'' (e.g. \cite{Santillan2000}). Thus, any assessment of the Parker Instability in a numerical model must address whether the instability can be resolved in length and time, as well as considering the input physics, the main sensitivities being the equation of state of the gas and the transport model for the cosmic rays.

Since Parker's original work, further analysis on the Parker Instability has been completed by a number of different papers. \cite{ryuparker2003} assumed finite diffusion parallel and perpendicular to the magnetic field and found that the parallel diffusion lowered the growth rate of the instability, while a small perpendicular diffusion had no noticeable effect. \cite{kuwabaraparker2004} and \cite{kuwabaraparker2006} also assumed finite diffusion parallel to the magnetic field and found that the growth rate decreased. This is due to the phase shift and reduction in magnitude of the cosmic ray pressure perturbation in the presence of diffusion.

In this paper, we  assess the effects of cosmic ray transport on the Parker Instability by considering two of the main models, self confinement and extrinsic turbulence \citep{zweibelreview2017}. In the self-confinement model, the cosmic rays excite Alfv\'{e}n waves through the streaming instability if their bulk velocity, or anisotropy, exceeds the Alfv\'{e}n speed $v_A$ \citep{Kulsrud1969}. These waves scatter the cosmic rays, with the point of marginal stability being cosmic ray isotropy in the frame of the waves. These waves transfer this energy and momentum to the thermal gas \citep{zweibelreview2017}. Another cosmic ray transport model is the model of ``extrinsic turbulence." In this model, the cosmic rays are driven to isotropy by scattering off of waves that are a result of a turbulent cascade, instead of being generated by the cosmic rays themselves. In this model, cosmic rays transfer momentum to the thermal gas through their pressure gradient, but do not heat it. Although both models can
accommodate diffusion, we neglect it in this work.

We discuss four different cases to obtain our eventual conclusions on the effect of cosmic ray transport on stratified systems. After presenting the basic equations in \S\ref{sec:setup}, we review the overstability of acoustic waves in the presence of cosmic ray streaming \citep{begelmanacoustic1994} in \S\ref{subsec:acoustic}. Then, we move into three different versions of the Parker Instability. In \S\S\ref{subsec:classic}, \ref{subsec:modified}, and \ref{subsec:streaming}
respectively, we reproduce the original result, then modify the compressibility of the cosmic rays and then finally add in cosmic ray streaming. In \S\ref{sec:physical} we compare the growth rates and domains of Parker Instability in the three cases (\S\ref{subsec:effect}), interpret the results
by discussing the positive and negative work terms in the growing modes (\S\ref{subsec:work}), and compare the 
stability criteria derived in two versus three dimensions. In \S \ref{sec:summary} we summarize the
main results and conclusions.

\section{Setup of the Problem} \label{sec:setup}
We assume a stratified system where the equilibrum cosmic ray pressure $P_c$, thermal gas pressure $P_g$, thermal gas density $\rho$, magnetic field $\mathbf{B}=B\hat{x}$, and gravitational acceleration $\mathbf{g}=-g\hat{z}$ are all functions of $z$ alone. The general linearized equations for the perturbed quantities are \citep{breitschwerdtwinds1991}:
\begin{gather}
\label{eq:contin}
\frac{\partial\delta\rho}{\partial t} = -\mathbf{\delta u} \cdot \del\rho - \rho\del \cdot \mathbf{\delta u} 
\\
\label{eq:momconsv}
\rho\frac{\partial\mathbf{\delta u}}{\partial t} = -\del(\delta P_c + \delta P_g) + \frac{\mathbf{J}\times\delta\mathbf{B}}{c} + \frac{\delta\mathbf{J} \times \mathbf{B}}{c} + \delta\rho\mathbf{g} 
\\
\frac{\partial\delta\mathbf{B}}{\partial t} = \del \times (\delta\mathbf{u} \times \mathbf{B}) 
\end{gather}
\begin{gather}
\begin{split}
\frac{\partial\delta P_c}{\partial t} + \mathbf{u}_A\cdot\del\delta P_c = -(\mathbf{\delta u} + \mathbf{\delta u}_A)\cdot\del P_c \\ 
-\gamma_c P_c \del \cdot (\mathbf{\delta u} + \mathbf{\delta u}_A)
\end{split}
\end{gather}
\begin{gather}
\begin{split} \label{eq:thermenergy}
\frac{\partial\delta P_g}{\partial t} = -\mathbf{\delta u} \cdot \del P_g - \gamma_g P_g\del \cdot \mathbf{\delta u} \\ 
-(\gamma_g - 1)(\mathbf{u}_A \cdot \del\delta P_c + \mathbf{\delta u}_A \cdot \del P_c)
\end{split}
\end{gather}
where $\mathbf{u}$ is the velocity, $\mathbf{u}_A$ is the standard Alfv\'{e}n velocity, $\mathbf{B}/\sqrt{4\pi\rho}$, and $\gamma_g$ and $\gamma_c$ are the thermal gas and cosmic ray adiabatic indices, respectively. Perturbations are denoted by $\delta$. Equations (\ref{eq:contin}) - (\ref{eq:thermenergy}) describe particle continuity, momentum conservation, magnetic induction, cosmic ray energy, and thermal energy, respectively. The energy equations include cosmic ray streaming terms, consistent with the self-confinement model. In our system, we had to assume there was a very weak horizontal pressure gradient which couples the cosmic rays to the background medium in equilibrium. We assume the cosmic rays are well-coupled and neglect diffusion. We also assume that any effects from thermal conduction are negligible.

The cosmic ray heating term is a result of the streaming cosmic rays depositing energy into the waves. In a steady state, these waves have constant amplitude, so they must transfer the energy gained from the cosmic rays into the background medium through a damping mechanism such as ion-neutral friction \citep{Kulsrud1969}, scattering by background magnetic inhomogeneity \citep{Farmer2004}, or a nonlinear mechanism \citep{KulsrudBook2005}. The Alfv\'{e}n speed scales this heating term and is strongest in the self-confinement model, while being zero in the extrinsic turbulence model due to the presence of oppositely directed waves which transfer energy to the cosmic rays rather than absorbing energy from them (assuming the turbulence is balanced).

It is worth noting that the velocity space anisotropy required to excite the streaming instability generally requires a cosmic ray pressure gradient parallel to the magnetic field. Thus, there is a slight inconsistency in taking an equilibrium model which is homogeneous in the parallel direction. We resolve this by assuming, as \cite{druryacoustic1986} and subsequent papers  did, that there is a parallel cosmic ray pressure  gradient with scale $L_{\parallel}$ and restricting our analysis to perturbations with parallel wavenumber $k_{\parallel}\gg L_{\parallel}^{-1}$. This suggests a corresponding limiting perturbation amplitude such that the linear analysis is valid: $\delta P_c/P_c < (k_{\parallel}L_{\parallel})^{-1}$. Above this amplitude, the streaming instability would excite waves moving in the opposite direction. This nonlinear effect will be included in future work.

In order to introduce notation and as a baseline for the rest of this paper, we first revisit the overstability of acoustic waves in a one-dimensional system due to cosmic rays, first analyzed by \cite{druryacoustic1986} and \cite{begelmanacoustic1994}. The acoustic case acts in some respects as a simplified form of the Parker Instability with streaming with which we can make comparisons later in the paper.

Then, we take our stratified system and put it in two dimensions for three different cases, all of which yield the Parker Instability. We  first reproduce the results of \cite{parkerinstability1966}, then modify his original case by taking $\gamma_c=4/3$. Lastly, we take this modified case and insert cosmic ray streaming. By examining all four of these cases, we can compare the results of each and reach conclusions about what is causing instability.

In all four cases, we follow the standard approach of reducing the set of linear partial differential equations for the state variables $(\delta\rho,\delta P_g,\delta P_c, \mathbf{\delta u},\mathbf{\delta B})$ to a set of linear algebraic equations and deriving a dispersion relation by setting the determinant of the system equal to zero. The dispersion relations for the acoustic case and classic Parker case are given in the main text. The dispersion relations for the modified Parker cases are given in the Appendix because of their unwieldy length.

For simplicity and to facilitate comparison with nonlinear numerical simulations, we work in two dimensions here. As mentioned in \S\ref{sec:intro}, the most
unstable modes of a stratified atmosphere with a horizontal magnetic field occur in the limit of infinitely short horizontal wavenumber perpendicular to $\mathbf{B}$. By carrying out the analysis in 2D, we are likely to bracket the degree of instability found in a 3D numerical simulation. In \S\ref{subsec:2D3D}, we compare the results of our 2D streaming case with the results from the 3D streaming case and explain the differences between the two models. 

\section{Results}\label{sec:results} 
\subsection{Overstability of Acoustic Waves}\label{subsec:acoustic}
Here, we  recover the dispersion relation for short wavelength, acoustic waves propagating parallel to $\mathbf{B}$ \citep{begelmanacoustic1994}. We drop all derivatives of background quantities and take $\mathbf{B}=B\hat{x}$. Thus, eqns. (\ref{eq:contin}) - (\ref{eq:thermenergy}) become:
\begin{gather}
\frac{\partial\delta\rho}{\partial t} = -\rho\frac{\partial\delta u_x}{\partial x} 
\end{gather}
\begin{gather}
\rho\frac{\partial\delta u_x}{\partial t} = -\frac{\partial}{\partial x}(\delta P_c + \delta P_g) 
\\
\frac{\partial\delta P_c}{\partial t} + u_A\frac{\partial\delta P_c}{\partial x} = - \gamma_c P_c \frac{\partial}{\partial x} (\delta u + \delta u_A)
\\
\frac{\partial\delta P_g}{\partial t} = - \gamma_g P_g \frac{\partial\delta u_x}{\partial x} - (\gamma_g - 1)u_A \frac{\partial\delta P_c}{\partial x}
\end{gather}
We also use:
\begin{equation}\label{eq:deltauA}
\delta u_A = -u_A \frac{\delta\rho}{2\rho}
\end{equation}
We assume that these perturbed quantities depend on $x$ and $t$ as $e^{i(kx-\omega t)}$. Using eqn. (\ref{eq:deltauA}) with this assumption, we reduce our equations to:
\begin{gather}
\omega\delta\rho = k\rho\delta u_x
\\
\omega\rho\delta u_x = k(\delta P_g + \delta P_c)
\\
(\omega - ku_A)\delta P_c = k\gamma_c P_c(u_x-u_A\frac{\delta\rho}{2\rho})
\\
\omega\delta P_g = k\gamma_g P_g u_x + (\gamma_g - 1)k u_A \delta P_c
\end{gather}
The dispersion relation resulting from this linear system is:
\begin{equation}\label{eq:acousticdisp}
\begin{split}
\omega(\omega-k u_A)&(\omega^2-k^2 a_g^2) \\
-k^2 a_c^2(\omega - k u_A/2)&(\omega + (\gamma_g-1)k u_A) = 0
\end{split}
\end{equation}
where we have followed \cite{begelmanacoustic1994} in defining $a_g^2 = \gamma_g P_g / \rho$ and $a_c^2 = \gamma_c P_c / \rho$, the squared gas and cosmic ray sound-speeds, respectively. Equation
(\ref{eq:acousticdisp})
 agrees with the result of \cite{begelmanacoustic1994}.

\begin{figure}[t!]
\plotone{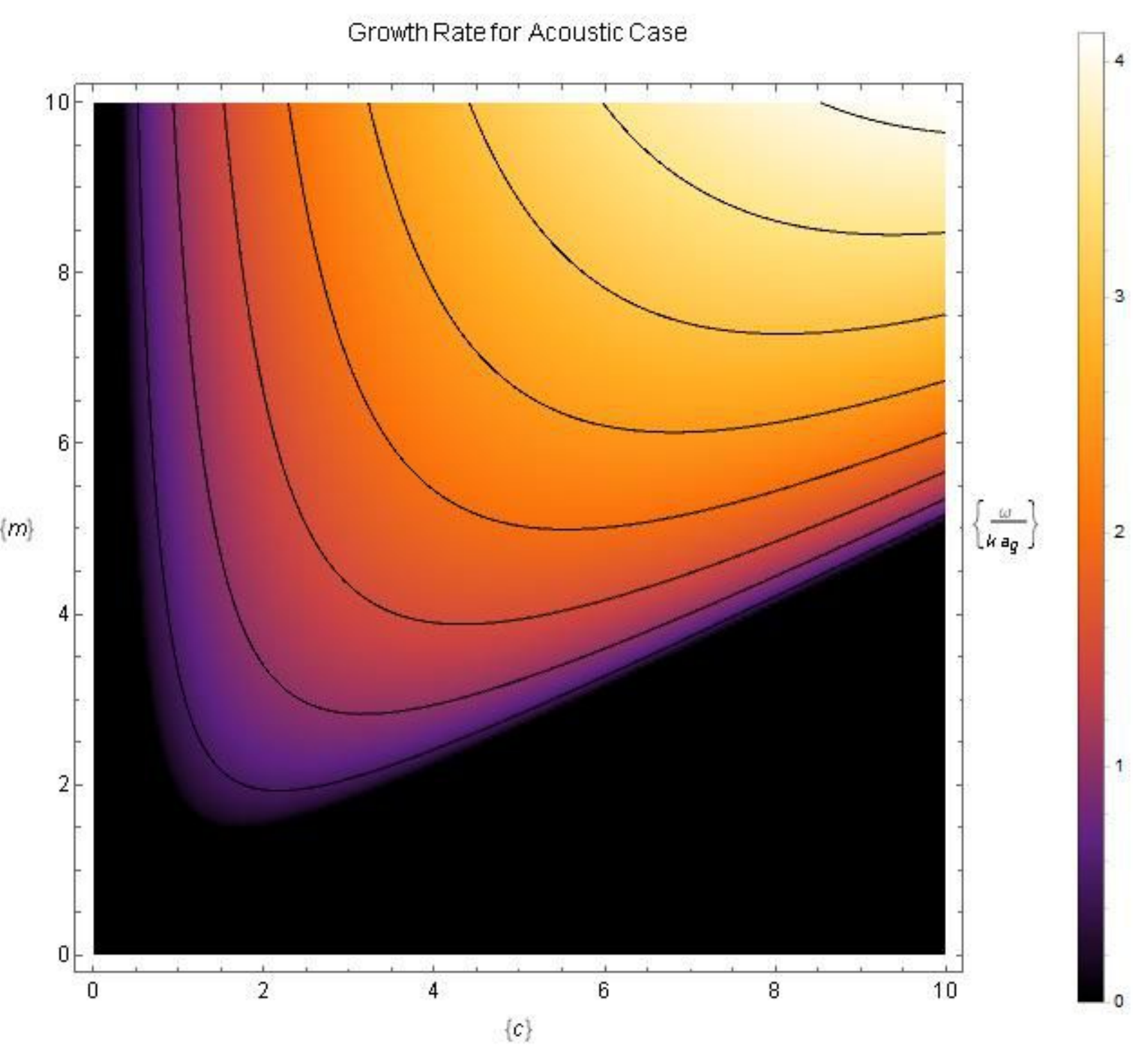}
\caption{The value of $c$ appears to have a larger effect on stability as its growth causes the system to become stable as we can see for values like $m=3$. The behavior is different for $m$ as once it becomes unstable, increasing $m$ just increases the growth rate. 
\label{fig:acousticContour}}
\end{figure}

Now, we wish to determine the conditions under which eqn. (\ref{eq:acousticdisp}) predicts instability and see how the instability grows as we change different parameters in the system. We follow the definitions used by \cite{begelmanacoustic1994}: 
\begin{equation}\label{eq:mc}
m \equiv \frac{u_A}{a_g}, \qquad c \equiv \frac{a_c}{a_g}, \qquad \hat{\omega}_{ac}\equiv\frac{\omega}{ka_g}
\end{equation}
Running a parameter study that iterates over these quantities, we make the contour plot found in Figure \ref{fig:acousticContour} showing the dimensionless growth rate for different values of $m$ and $c$.

As discussed in \cite{begelmanacoustic1994}, the overstability of acoustic waves is due to cosmic ray heating. No matter how small $c$ is, a sufficiently large value of $m$ will yield instability through heating. On the other hand, fixing $m$ and letting $c$ increase yields stable waves driven by cosmic ray pressure as a restoring force for sufficiently large $c$.

\subsection{Classic Parker}\label{subsec:classic}
We now reproduce the stability derivation of \cite{parkerinstability1966}. In this and the two following subsections, we assume that in equilibrium the thermal gas is isothermal ($P_g/\rho$ is constant) and that the ratios of cosmic ray to thermal pressure and magnetic to thermal pressure are constant as well. Following Parker, we assume that the cosmic rays behave as a $\gamma_c = 0$ fluid while affecting the stratification through their pressure gradient. In his original treatment, Parker neglected any cosmic ray streaming terms. Therefore, for this case, we drop any terms associated with $u_A$ since they represent our cosmic rays streaming at the Alfv\'{e}n speed. We work in two dimensions, assuming that perturbed quantities depend on $x$ and $z$. 

We define a scale height as: 
\begin{equation}
H = \frac{H_0}{q}
\end{equation} 
where $H_0 = a_g^2 / g$ is a fiducial scale height and:
\begin{equation}\label{eq:q}
q\equiv\frac{\gamma_g}{1+\frac{\gamma_g}{\gamma_c}c^2+\frac{\gamma_g}{2}m^2}
\end{equation}
with $m$ and $c$ as defined in eqn. (\ref{eq:mc}). With these assumptions and
definitions, $\rho$, $P_c$, $P_g$, and $B^2$ all depend on $z$ as $e^{-z/H}$.

\begin{figure}[t!]
\gridline{\fig{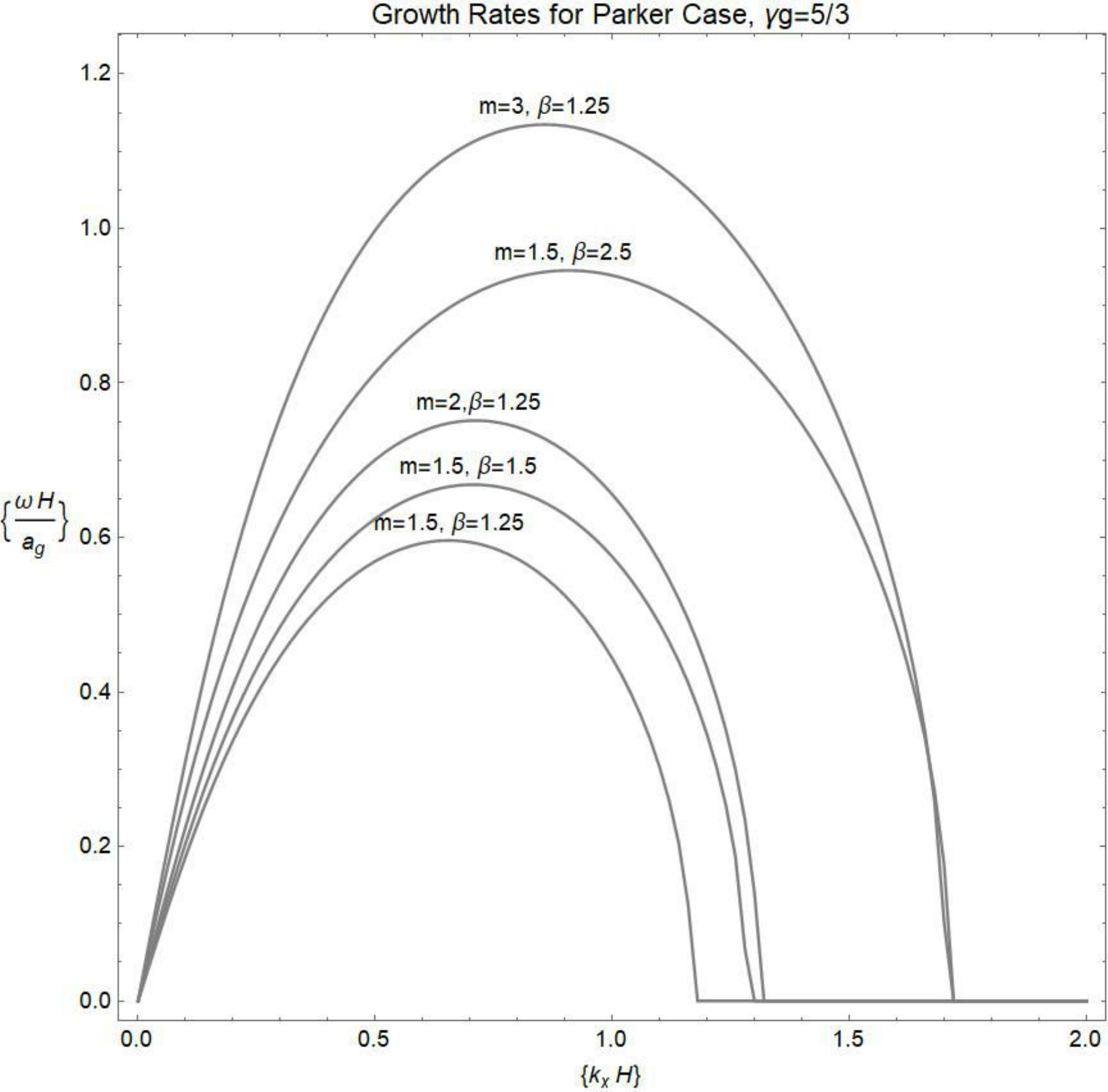}{0.37\textwidth}{(a)}
		  }
\gridline{\fig{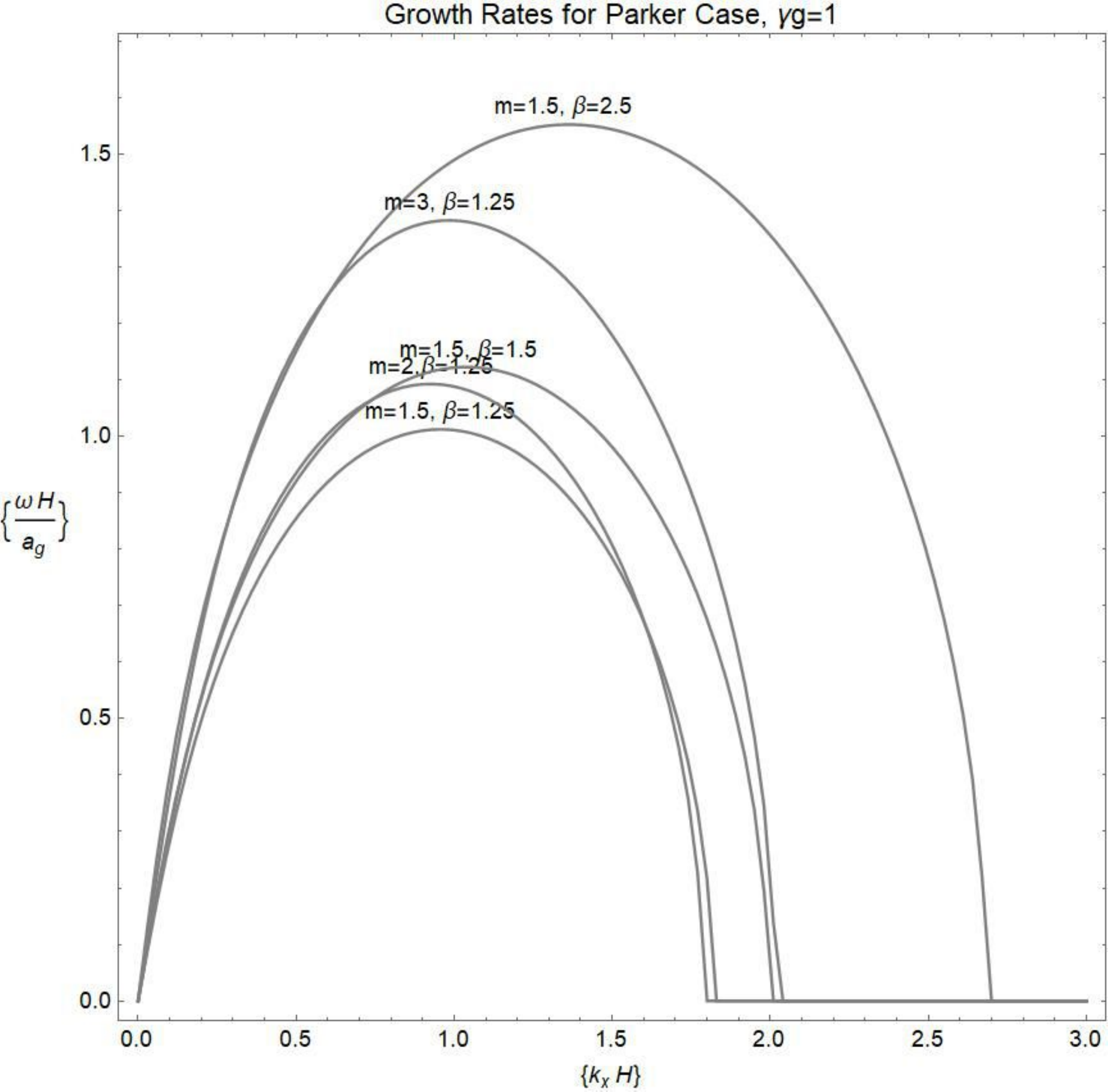}{0.37\textwidth}{(b)}
		  }        
\caption{Plots (a) and (b) show the growth rate for Classic Parker for different values of $m$ and $\beta$ with $\gamma_g=5/3$ (a) and $\gamma_g=1$ (b). As we increase both $m$ and $\beta$, we note that the unstable modes occur over a larger range of $\hat{k_x}$ values and the growth rate peaks at a greater value. We also note that as $\gamma_g$ gets smaller from (a) to (b), the growth rate gets slightly larger and the system is unstable over a larger range of $\hat{k_x}$ values.
\label{fig:classicParker}}
\end{figure}

We nondimensionalize the perturbation quantities as follows:
\begin{gather}\label{eq:dimendefn}
\begin{split}
\hat{\omega} = \frac{\omega H}{a_g}, \qquad \hat{k} = kH,\\
\hat{\delta\rho} = \frac{\delta\rho}{\rho}, \qquad \hat{\delta B} = \frac{\delta B}{B},
\end{split}
\end{gather}
\begin{gather}
\begin{split}
\hat{\delta P_c} = \frac{\delta P_c}{P_c}, \qquad \hat{\delta P_g} = \frac{\delta P_g}{P_g}, \\
\alpha = \frac{\gamma_gm^2}{2}, \qquad \beta = \frac{P_c}{P_g}.
\end{split}
\end{gather}
The $\alpha$ and $\beta$ parameters are the same ones introduced by Parker. The perturbed quantities have harmonic dependence on $x$ and $t$ while the $z$ dependences are:
\begin{equation}
\delta u \propto e^{(ik_z z+\frac{z}{2H})}, \qquad \delta P \propto  e^{(ik_z z-\frac{z}{2H})}, \qquad \delta B \propto e^{ik_z z}
\end{equation}
Using these definitions along with our previous definition of $m$ ($c=0$ here), our linearized equations are:
\begin{gather}
i\hat{\omega}\delta\hat{\rho} - i\hat{k_x}\delta\hat{u_x} - (i\hat{k_z}-\frac{1}{2})\delta\hat{u_z} = 0 \\
i\gamma_g\hat{\omega}\hat{\delta u_x}-i\hat{k_x}(\beta\hat{\delta P_c}+\hat{\delta P_g})-\alpha\hat{\delta B_z} = 0 \\
\begin{split}
i\gamma_g\hat{\omega}\hat{\delta u_z}-(i\hat{k_z}-\frac{1}{2})(\beta\hat{\delta P_c}+\hat{\delta P_g}+2\alpha\hat{\delta B_x}) \\
+2i\alpha\hat{k_x}\hat{\delta B_z}-\frac{\gamma_g}{q}\hat{\delta\rho}= 0 
\end{split} \\
\hat{\omega}\hat{\delta B_x}-\hat{k_z}\hat{\delta u_z} = 0 \\
\hat{\omega}\hat{\delta B_z}+\hat{k_x}\hat{\delta u_z} = 0 \\
i\hat{\omega}\hat{\delta P_c}+\hat{\delta u_z} = 0 \\
i\hat{\omega}\hat{\delta P_g}-i\gamma_g\hat{k_x}\hat{\delta u_x}+(1-\frac{\gamma_g}{2}-i\gamma_g\hat{k_z})\hat{\delta u_z} = 0
\end{gather}
The resulting dispersion relation is:
\begin{equation}\label{eq:classicparkdisp}
\begin{split}
\gamma_g^2\hat{\omega}^4-\hat{\omega}^2(\gamma_g^2+2\alpha&\gamma_g)(\hat{k_x}^2+\hat{k_z}^2+\frac{1}{4})+\hat{k_x}^2\Big(2\alpha\gamma_g(\hat{k_x}^2+\hat{k_z}^2) \\
-2(\alpha+\beta+\frac{1}{2})&-(\alpha+\beta)^2+\gamma_g(\frac{3\alpha}{2}+\beta+1)\Big) = 0
\end{split}
\end{equation}
which matches the result from \cite{parkerinstability1966}. We then let $\gamma_g=5/3$, chose values for our two parameters $m$ and $\beta$, and solved the dispersion relation by running over values of $\hat{k_x}$ and $\hat{k_z}$. In Figure \ref{fig:classicParker}, we have plotted the maximum growth rate vs. $\hat{k_x}$ for different values of $m$ and $\beta$, while $\hat{k_z}=0$, which maximizes the instability with respect to $\hat{k_z}$.
From Figure \ref{fig:classicParker},we can observe that an increase in the value of $m$ leads to a larger maximum growth rate and causes it to occur at a larger wavenumber. Increasing $\beta$ creates the same effect.

Another interesting comparison in this Classic Parker case is to see how the instability changes as $\gamma_g$ changes from $\gamma_g=5/3$ to $\gamma_g=1$. It is argued in \cite{parkerinstability1966} that $\gamma_g = 1$ is the correct value due to the radiative efficiency of the gas and presumed constancy or even decrease in the turbulent velocity dispersion with gas density. In Figure \ref{fig:classicParker}, we have plotted the
growth rates for $\gamma_g=5/3$ and $\gamma_g=1$ to see the difference. As we decrease the value of $\gamma_g$, we see that the system becomes more unstable, as it takes less work to compress the gas into the  valleys of the undular perturbed magnetic field, an effect also seen in \cite{zweibelparker1975}. This begins to give us an indication of the effect that changing the value of $\gamma_c$ will have on the Modified Parker case.

\subsection{Modified Parker}\label{subsec:modified}
We now move to the Modified Parker case that was first mentioned by \cite{zweibelparker1975} and \cite{boettcherEDIG2016}. In this case, the major difference is that instead of letting $\gamma_c=0$ as
in Parker's original treatment, we set $\gamma_c=4/3$, which is consistent with the extrinsic turbulence model of cosmic ray transport. In this case, instead of using $\alpha$ and $\beta$ in our equations, we return to using $m$ and $c$ defined in eqn. (\ref{eq:mc}). Since $\gamma_c$ is no longer zero, 
we also redefine our dimensionless quantities for $\delta P_g$ and $\delta P_c$ where:
\begin{equation}\label{eq:newpertpress}
\hat{\delta P_c}=\frac{\delta P_c}{\gamma_c P_c} \qquad 
\hat{\delta P_g}=\frac{\delta P_g}{\gamma_g P_g}
\end{equation}
Our other dimensionless quantity definitions are as in eqn. (\ref{eq:dimendefn}). Furthermore, we continue to assume the same exponential dependence on $x$, $z$, and $t$ as we used for Classic Parker. Again ignoring the streaming terms, the linearized equations are:
\begin{gather}
\label{eq:modparkcont}
i\hat{\omega}\hat{\delta\rho}-i\hat{k_x}\hat{\delta u_x}-(i\hat{k_z}-\frac{1}{2})\hat{\delta u_z} = 0 \\
i\hat{\omega}\hat{\delta u_x}-i\hat{k_x}(c^2\hat{\delta P_c}+\hat{\delta P_g})-\frac{m^2}{2}\hat{\delta B_z} = 0 \\
\begin{split}
i\hat{\omega}\hat{\delta u_z}-(i\hat{k_z}-\frac{1}{2})(c^2\hat{\delta P_c}+\hat{\delta P_g}+m^2\hat{\delta B_x}) \\
+im^2\hat{k_x}\hat{\delta B_z}-\frac{1}{q}\hat{\delta\rho} = 0 
\end{split} \\
\hat{\omega}\hat{\delta B_x}-\hat{k_z}\hat{\delta u_z} = 0 \\
\hat{\omega}\hat{\delta B_z}+\hat{k_x}\hat{\delta u_z} = 0 \\
i\hat{\omega}\hat{\delta P_c}-i\hat{k_x}\hat{\delta u_x}+(\frac{1}{\gamma_c}-i\hat{k_z}-\frac{1}{2})\hat{\delta u_z} = 0 \\
\label{eq:modparkgas}
i\hat{\omega}\hat{\delta P_g}-i\hat{k_x}\hat{\delta u_x}+(\frac{1}{\gamma_g}-i\hat{k_z}-\frac{1}{2})\hat{\delta u_z} = 0
\end{gather}

As with the Classic Parker case, we choose a value for $\gamma_g$, select values for $m$ and $c$, and solve the eigenvalue problem corresponding to the linearized equations
by running over values of $\hat{k_x}$ and $\hat{k_z}$. We have provided the dispersion relation for this system as eqn. \ref{eq:modparkerdisp} in Appendix \ref{sec:modparkdisp}.

In Figure \ref{fig:modifiedParker}, we again plot the maximum growth rate vs. $k_x$ for $\gamma_g = 5/3$ (a) and
$\gamma_g=1$ (b), the same values of $m$ and $\hat{k_z}=0$. We choose the values of $c$ that would correspond to the same value of $\beta$ for $\gamma_c=4/3$.
\begin{figure}[t!]
\gridline{\fig{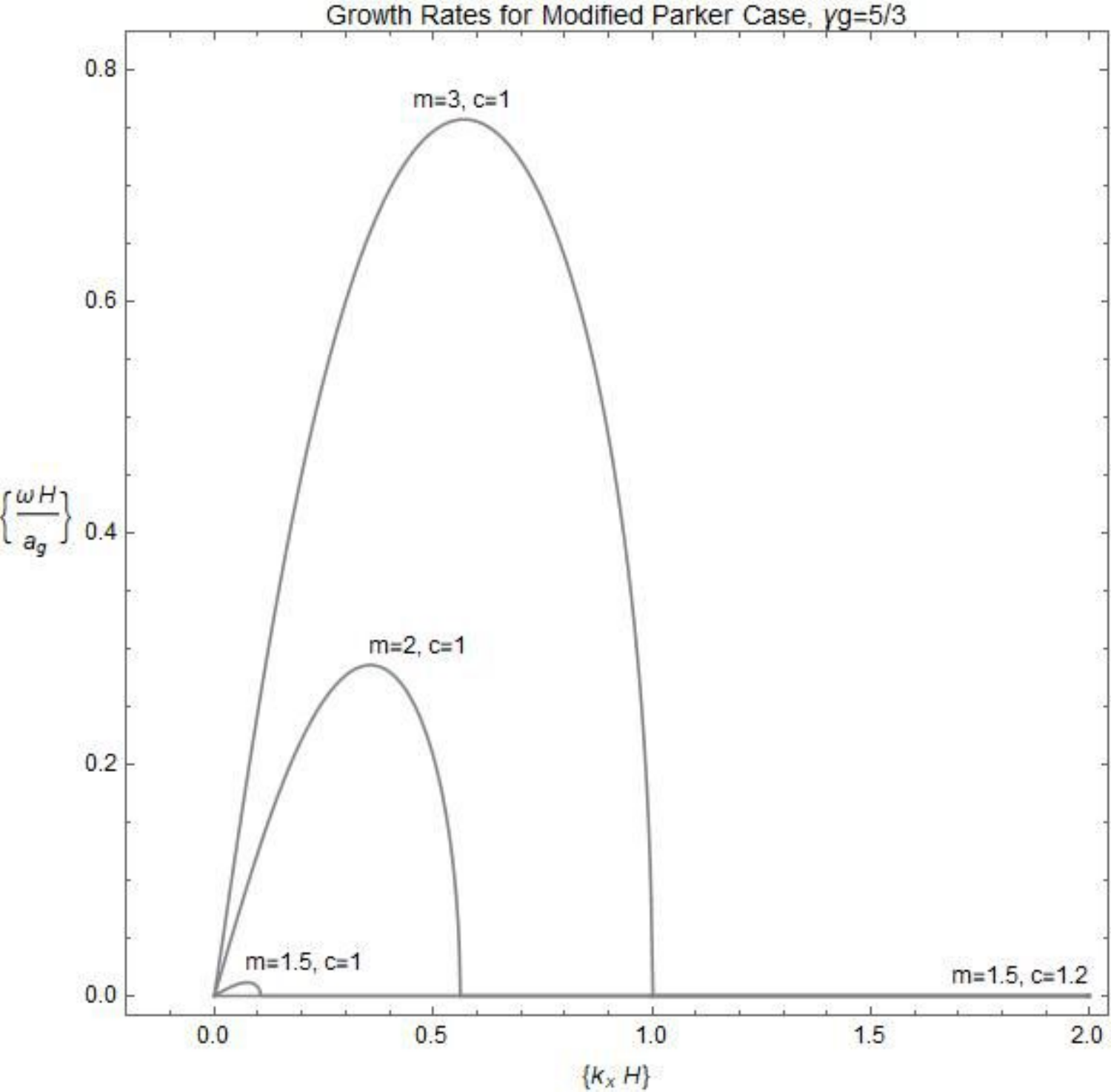}{0.37\textwidth}{(a)}
		  }
\gridline{\fig{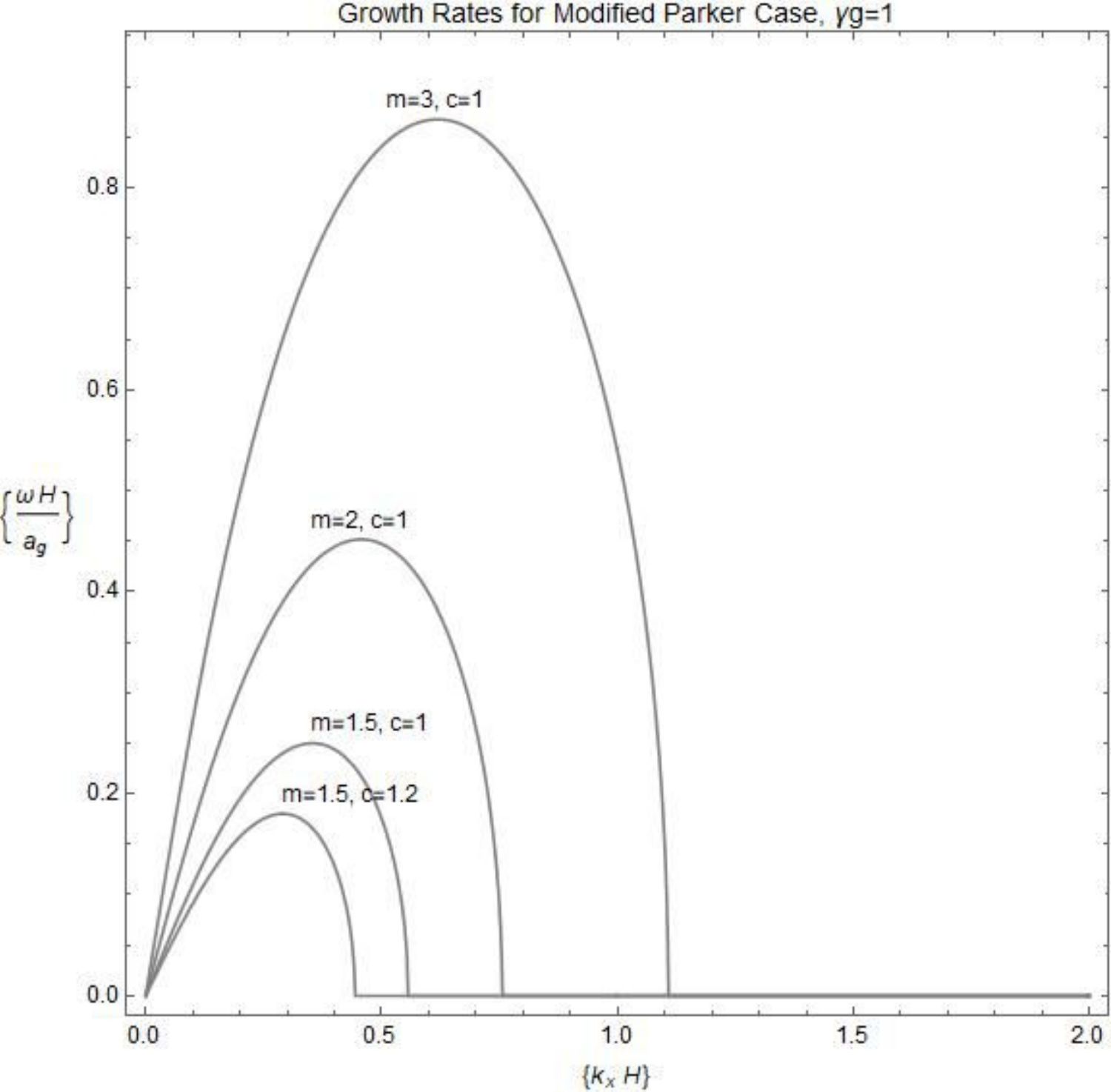}{0.37\textwidth}{(b)}
		  }        
\caption{Plots (a) and (b) show the growth rate for Modified Parker for different values of $m$ and $c$ with $\gamma_g=5/3$ (a) and $\gamma_g=1$ (b). We have neglected the $c=2$ contour here as when $c$ was increased to that value, the instability disappears. As we increase the value of $m$, however, the instability occurs over a larger range of wavelengths and peaks at a greater value, as we saw for Classic Parker, while the growth rates are slightly larger for the lower value of $\gamma_g$.}
\label{fig:modifiedParker}
\end{figure}
We have omitted the $c=2$ contour corresponding to the contour in Figure \ref{fig:classicParker} where we increased $c$ to $c=2$ and kept $m$ constant. The reason for this is that the $c=2$ case had no unstable modes. This is largely due to the fact that with the change of $\gamma_c$ from $0$ to $4/3$, the cosmic rays act as a stabilizing force in this system, as evidenced by the lower growth rate seen when $c$ is increased from $c=1$ to $c=1.2$. The increase in the cosmic rays' adiabatic index means that they are harder to compress in our fluid model, which offsets the potential energy liberated by moving thermal gas into the magnetic valleys. As expected, however, the growth rates are higher for the softer thermal gas equation of state.

\subsection{Modified Parker with Cosmic Ray Streaming}\label{subsec:streaming}
For our final model, we now insert cosmic ray streaming into our Modified Parker Case. Therefore, in our linearized equations, we now add back in the streaming terms from the self-confinement model of cosmic ray transport which are associated with the Alfv\'{e}n speed. We keep the same definitions for our dimensionless quantities from eqns. (\ref{eq:dimendefn}) and (\ref{eq:newpertpress}) and so our linearized equations are:
\begin{gather}
\label{eq:parkstreamcont}
i\hat{\omega}\hat{\delta\rho}-i\hat{k_x}\hat{\delta u_x}-(i\hat{k_z}-\frac{1}{2})\hat{\delta u_z} = 0 \\
i\hat{\omega}\hat{\delta u_x}-i\hat{k_x}(c^2\hat{\delta P_c}+\hat{\delta P_g})-\frac{m^2}{2}\hat{\delta B_z} = 0 \\
\begin{split}
i\hat{\omega}\hat{\delta u_z}-(i\hat{k_z}-\frac{1}{2})(c^2\hat{\delta P_c}+\hat{\delta P_g}+m^2\hat{\delta B_x}) \\
+im^2\hat{k_x}\hat{\delta B_z}-\frac{1}{q}\hat{\delta\rho} = 0
\end{split} \\
\hat{\omega}\hat{\delta B_x}-\hat{k_z}\hat{\delta u_z} = 0 \\
\hat{\omega}\hat{\delta B_z}+\hat{k_x}\hat{\delta u_z} = 0 \\
\begin{split}
(i\hat{\omega}-i\hat{k_x}m)\hat{\delta P_c}+\frac{i\hat{k_x}m}{2}\hat{\delta\rho}-i\hat{k_x}(\hat{\delta u_x}+m\hat{\delta B_x}) \\
+(\frac{1}{\gamma_c}-i\hat{k_z}-\frac{1}{2})(\hat{\delta u_x}+m\hat{\delta B_z}) = 0 
\end{split} \\
\begin{split}
\label{eq:parkstreamgas}
i\hat{\omega}\hat{\delta P_g}-i\hat{k_x}(\hat{\delta u_x}+mc^2(\gamma_g-1)\hat{\delta P_c}) \\
+(\frac{1}{\gamma_g}-i\hat{k_z}-\frac{1}{2})\hat{\delta u_z}+\frac{mc^2}{\gamma_c}(\gamma_g-1)\hat{\delta B_z} = 0
\end{split}
\end{gather}

As with the Modified Parker case, we consider both $\gamma_g = 5/3$ and $\gamma_g = 1$ and set $\gamma_c=4/3$. We again select values for $m$ and $c$ and numerically solve the dispersion relation by running over values of $\hat{k_x}$ and $\hat{k_z}$. The dispersion relation for this system is also provided by eqn. \ref{eq:parkerstreamdisp} in Appendix \ref{sec:modparkdisp}.

\begin{figure}[t!]
\gridline{\fig{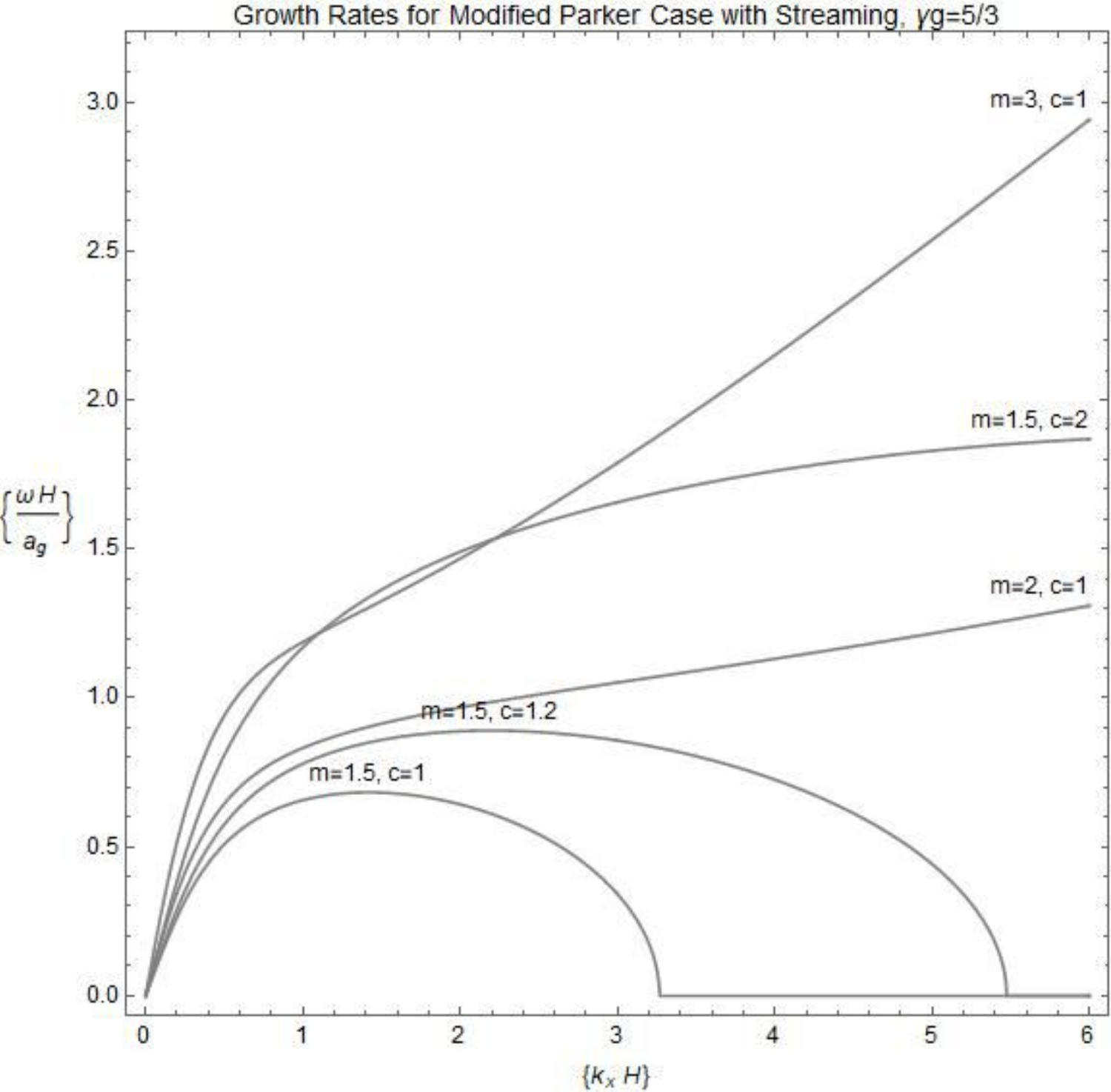}{0.37\textwidth}{(a)}
		  }
\gridline{\fig{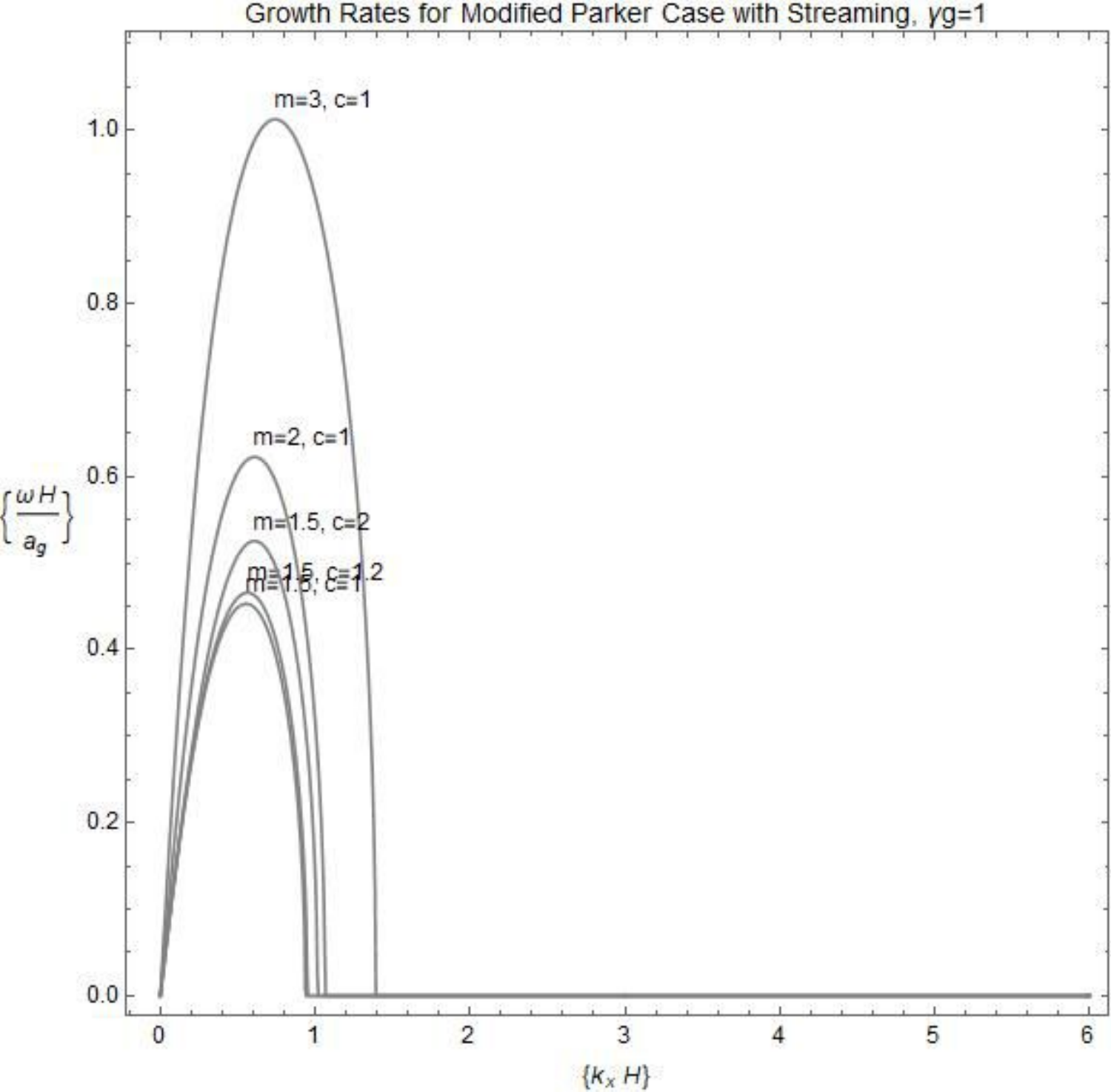}{0.37\textwidth}{(b)}
		  }
\caption{Plots showing the growth rate for Modified Parker with Cosmic Ray Streaming for
$\gamma_g = 5/3$ (a) and $\gamma_g = 1$ (b), and different values of $m$ and $c$. In (a), as we increase both $m$ and $c$, the instability occurs over a larger domain and peaks at a larger value, as we previously saw for the Classic Parker Case. Note that this differs from the Modified Parker Case where increasing the value of $c$ led to no unstable modes. For (b), the cosmic rays are advected relative to the thermal gas but do not heat it. This results in slightly larger growth rates than
for Modified Parker, but is otherwise similar to that case.
\label{fig:parkerStream}}
\end{figure}

We plot the maximum growth rate vs. $\hat{k_x}$ for this case in Figure \ref{fig:parkerStream}, using the same values of $m$, $c$, and $\hat{k_z}$ as before. 

As with the previous two cases, increasing the ratio of the Alfv\'{e}n speed to the gas sound-speed causes the system to become more unstable. However, in this final case, for the values of $m$ and $c$ that we have used throughout the three cases and $\gamma_g = 5/3$, the system never reaches a stability boundary as it does in the other two cases. In contrast, setting $\gamma_g = 1$ eliminates cosmic ray heating and produces results which are very similar to the Modified Parker case. This strongly implicates cosmic ray heating as the source of enhanced instability.

\section{Physical Interpretation}\label{sec:physical}
Having outlined each case separately, we now compare their results and aim to determine what is causing the instability. First, we analyze the three contour plots for each case and compare the domain of instability as well as the growth rate of these instabilities. Then, we determine the work contributions from the different components in each of our systems to see what is stabilizing the system and what is creating instability. Finally, we compare the two dimensional streaming case to its three dimensional counterpart.

\subsection{The Effect of Streaming}\label{subsec:effect}
\begin{figure}
\gridline{\fig{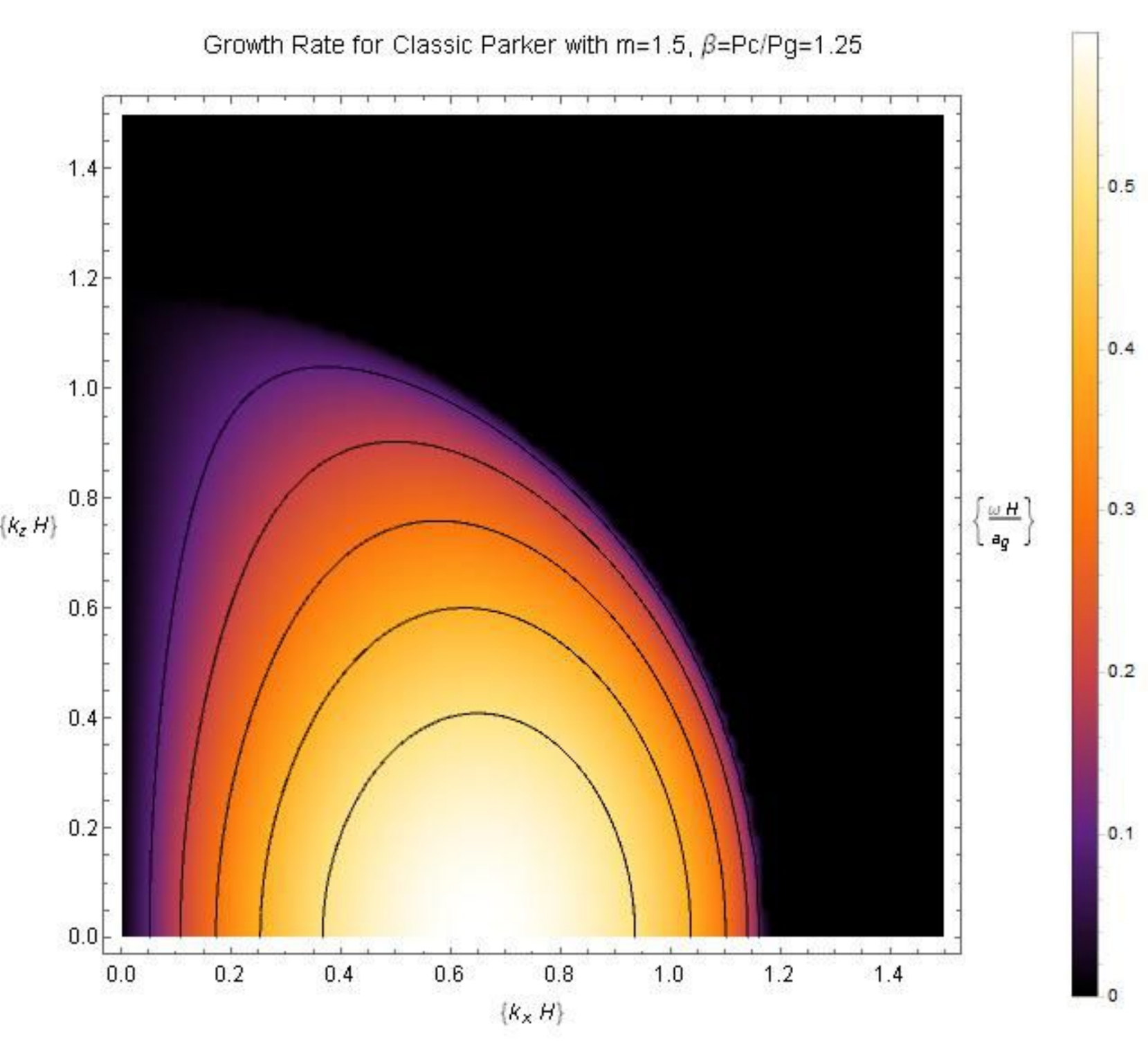}{0.37\textwidth}{(a)}
		  }
\gridline{\fig{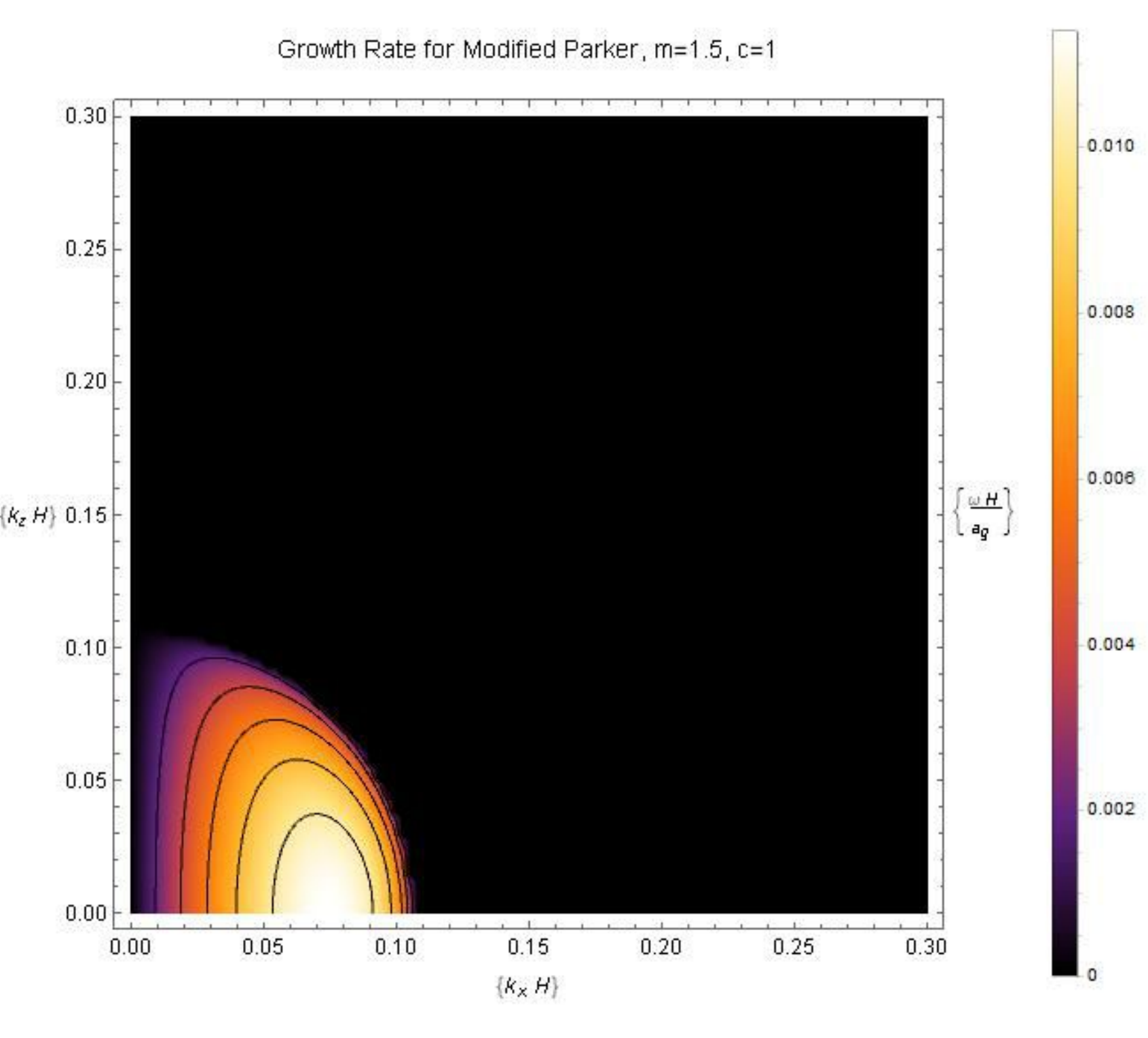}{0.37\textwidth}{(b)}
          }
\gridline{\fig{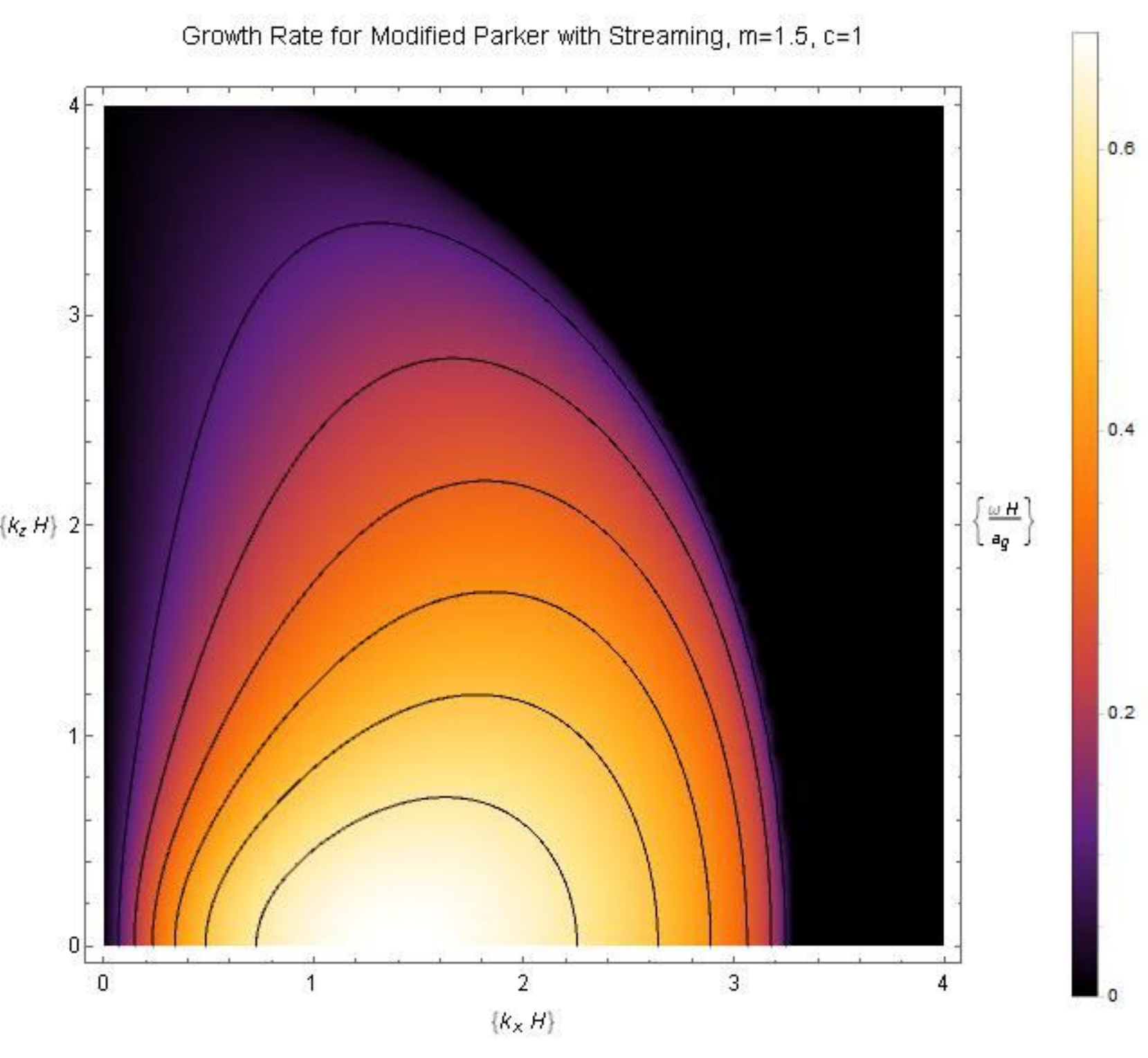}{0.37\textwidth}{(c)}
		  }
\caption{The contour plots of growth rate for the three Parker cases we have investigated in this paper. Note the different domains of $\hat{k_x}$ and $\hat{k_z}$ in the graphs as well as the different ranges for $\hat{\omega}$ in the bar legend. In all three cases, $\gamma_g=5/3$.}
\label{fig:contourPlots}
\end{figure}

The three contour plots of growth rate from the cases we have outlined in this paper are shown in Figure \ref{fig:contourPlots}. Between the three plots, we have used three different bar legends which show the ranges of the dimensionless growth rate. We also have used different domains for the dimensionless $\hat{k_x}$ and $\hat{k_z}$ in order to more easily show the full range of unstable modes in the three cases. For all three plots, we have let $m=1.5$ and $\beta=1.25$ which means $c=1$ in the final two plots. 

If we just compare Classic Parker with the Modified Parker case, we see that changing $\gamma_c=0$ to $\gamma_c=4/3$ almost completely stabilizes the system. In fact, not only does the range of unstable modes significantly decrease but the growth rate decreases by a factor of about $500$ as well. As we increase the value of $\gamma_c$, the cosmic rays require more energy to be compressed. Therefore, this offsets gravitational potential energy released by gas sliding downward into the magnetic valleys, helping
to stabilize the system and explaining the large gap in growth rates and domain of instablility between the two cases.

If we compare the Modified Parker case then to the Modified Parker with Streaming, we see that the system becomes much more unstable. The maximum growth rate is larger by a factor of about $700$, the horizontal wavelength of the fastest growing mode decreases, and the range of instability is larger as well. Therefore, the instability grows more quickly and the maximum growth rate requires a finer grid to resolve numerically. 

Furthermore, even comparing the Modified Parker with Streaming to Classic Parker, we see that the maximum growth rate slightly increases and that the range of instability almost doubles. Therefore, the Modified Parker with Streaming is our most unstable case.

\begin{deluxetable*}{c|c|cCccc}[t!]
\scriptsize
\tablecaption{Unstable Mode Work Contributions
\label{tab:unstablework}}
\tablecolumns{7}
\tablenum{1}
\tablewidth{0pt}
\tablehead{
	\colhead{} & \colhead{Terms} & \colhead{Acoustic} & \colhead{Classic Parker} & 			\colhead{Modified Parker} & \colhead{Modified Parker with Streaming}
}
\startdata
Parameter Values &  & $c=3$; $m=1.9$ & $c=1$; $m=1.5$ & $c=1$; $m=1.5$ & $c=1$; $m=1.5$ \\
\hline
Gas Pressure & $-\delta \mathbf{u} \cdot \nabla\delta P_g$ & $+$ & $-$ & $-$ & $+$ \\
Cosmic Ray Pressure & $-\delta \mathbf{u} \cdot \nabla\delta P_c$ & $-$ & $+$ & $-$ & $-$ \\
Gravity & $- \delta u_z g\delta\rho$ & N/A & $+$ & $+$ & $-$ \\
Magnetic Pressure & $-\delta u_z \frac{\partial}{\partial z} \frac{B\delta B_x}{4\pi}$ & N/A & $-$ & $-$ & $-$ \\
Magnetic Buoyancy & $\delta u_x \frac{\delta B_z}{4\pi}\frac{dB}{dz}$ & N/A & $+$ & $+$ & $+$ \\
Magnetic Tension & $\delta u_z\frac{B}{4\pi}\frac{\partial\delta B_z}{\partial x}$ & N/A & $-$ & $-$ & $-$ \\
\enddata
\tablecomments{The ``$+$'' means that the system is contributing positive work and therefore is destabilizing the system. The ``$-$'' means that the system is contributing negative work and so is stabilizing the system. For the acoustic case, only the cosmic ray pressure and gas pressure could contribute to the work from their perturbations so the other quantities are not applicable for that case.}
\end{deluxetable*}

As a numerical example, we let $m=1.5$, $c=1$, $\gamma_g=5/3$, $\gamma_c=4/3$, $a_g=10$ km/s, $H=248$ pc, and $\hat{k_z}=0$ since the maximum always occurs on the $\hat{k_x}$ axis. For the Modified Parker case, we find the approximate maximum growth rate of $\hat{\omega_i}\approx0.011447$ occurs at about $\hat{k_x}\approx0.075$, which for no streaming is close to the instability threshold. This leads to a growth time of $t\approx2.12$ Gyrs. and a wavelength of $\lambda\approx20.8$ kpc. However, if we add in streaming for the same values, the maximum growth rate jumps to $\hat{\omega_i}\approx0.68251$ at $\hat{k_x}\approx1.4$. This translates to a growth time of $t\approx35.6$ Myrs. and a wavelength of $\lambda\approx1.11$ kpc. If we omit streaming but set $\gamma_g=1$, the approximate maximum growth rate is $\hat{\omega_i}\approx0.249967$ at $\hat{k_x}\approx0.35$. The growth time is then $t\approx97.1$ Myr. and the wavelength is $\lambda\approx4.45$ kpc.

We can compare these numbers with the parameter values used in the galaxy model from \cite{ruszkowskiwinds2017}. In their paper, they assumed that $\gamma_c=4/3$ and $\gamma_g=5/3$, but cooling was included, making the effective $\gamma_g$ smaller. Simulation results are plotted out to 500 Myr, and the highest spatial resolution corresponded to 195 pc. Based on our results, their puffed-up disk with no cosmic ray streaming may have been stabilized by the stiff cosmic ray equation of state. However, near threshold, the simulation would have needed to run $4\times$ as long to allow the instability to develop. With $\gamma_g=1$ or streaming inserted, the instability would have been only marginally spatially resolved (however, including streaming leads to development of a wind).

\subsection{Work Contributions}\label{subsec:work}
We now aim to determine what is responsible for stabilizing and destabilizing the system. In order to do this, we look at the momentum conservation equation (\ref{eq:momconsv}) for both the $x$ and $z$ directions for the four cases. If we take the scalar product with $\delta{\mathbf{u}}$ and time average, we can determine the total work done for the unstable modes of each case. Furthermore, in each of these cases, we can look at the contribution from each of the perturbation quantities separately to determine what quantities are stabilizing and destabilizing the system. Table \ref{tab:unstablework} shows the unstable mode work contributions. We chose the mode that was closest to the stability boundary to see how the mode behaved as it prepared to transition to stability.

For the unstable modes, the Classic Parker and Modified Parker work contributions are pretty similar except for the cosmic ray pressure contribution. It destabilizes Classic Parker while stabilizing Modified Parker which appears to be due to the change in $\gamma_c$. As we stated before, this change means more energy is required to compress the cosmic rays and so they stabilize the Modified Parker case. The Modified Parker with Streaming is quite different from these two cases, however. With streaming, the cosmic rays and gravity become stabilizing while the gas becomes destabilizing. 

Based on the numerical results around this stability boundary, for our chosen $m$ and $c$, the gas pressure contributes about $2.5\times$ the work done by the magnetic buoyancy and therefore is largely responsible for the instability of these modes near the stability boundary. 
Note that thermal gas pressure is also destabilizing in the acoustic case, where it is understood to be the result of heating the gas during its phase of compression (\cite{begelmanacoustic1994}). Therefore, we attribute the enhancement of the Parker Instability by cosmic ray streaming to an analogous effect. 

This conclusion is bolstered by comparison of plots \ref{fig:parkerStream}a and \ref{fig:parkerStream}b, which differ mainly in that cosmic ray heating is present in (a) but not in (b).
A similar analysis of the stable mode work contributions is consistent with these results, showing that the cosmic ray heating is destabilizing the system while terms like the magnetic pressure are keeping it stable.

\subsection{2D vs. 3D}\label{subsec:2D3D}
As we mentioned in \S\ref{sec:setup}, we assumed for our different cases of the Parker Instability that our perturbations only depended on two coordinates, $x$ and $z$. However, it is known from previous analysis \citep{newcombinstability1961,parkerinstability1966} that the most stringent instability criterion is obtained in three dimensions and holds in the limit of infinitely short wavelength in the horizontal direction perpendicular to $\mathbf{B}$. While we have no guarantee that this is also true in the presence of cosmic ray streaming, and direct numerical simulations cannot achieve this limit, it seems only prudent to consider it here. We hope that by doing so we can bracket the stability criterion that would hold for a 3D system simulated with finite numerical resolution.

Therefore, in this section, we aim to compare our 2D streaming case to the same system, but in 3D. For the 3D case, instead of $k_y\rightarrow 0$ as in the 2D case, we assume that $k_y\rightarrow\infty$. By assuming that $k_y\rightarrow\infty$, we follow the original treatment from \cite{newcombinstability1961}, 
which showed that the most unstable perturbations have 
a parallel wavenumber $k_x\rightarrow 0$, horizontal perpendicular wavenumber $k_y\rightarrow\infty$, and 
total pressure perturbation of approximately zero. This minimizes the stabilizing effects of magnetic tension and 
total pressure, which would exert Alfvenic and magnetosonic restoring forces.

For the 3D case, a similar analysis to the one we performed in 2D along with using the assumptions listed above leads to the dispersion relation:
\begin{equation}
\omega^2-k^2u_A^2 = -\frac{g}{\rho}\frac{d\rho}{dz}-\frac{(\omega^2-k^2u_A^2)\rho g^2}{\Gamma\Pi(\omega^2-k^2u_A^2)+\rho k^2u_A^4}
\end{equation}
where:
\begin{equation}
\Gamma\Pi = \rho u_A^2+\gamma_gP_g+\gamma_cP_c\frac{\omega-ku_A/2}{\omega-ku_A}(1+(\gamma_g-1)\frac{ku_A}{\omega})
\end{equation}
which allows us to plot its result against the 2D plot to get the limits on the 3D instability.

\begin{figure}[ht!]
\gridline{\fig {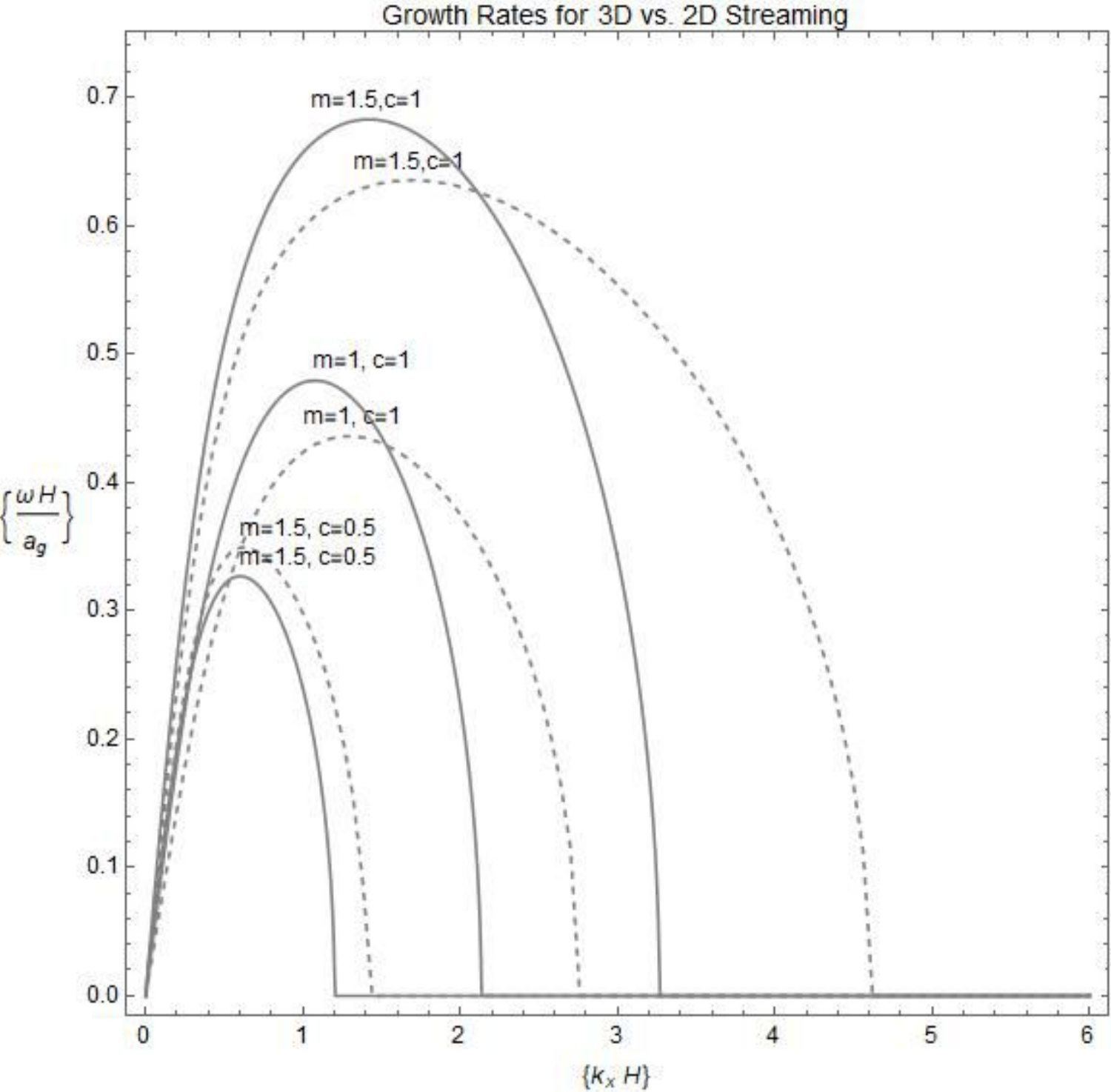}{0.37\textwidth}{(a)}
		 }
\gridline{\fig {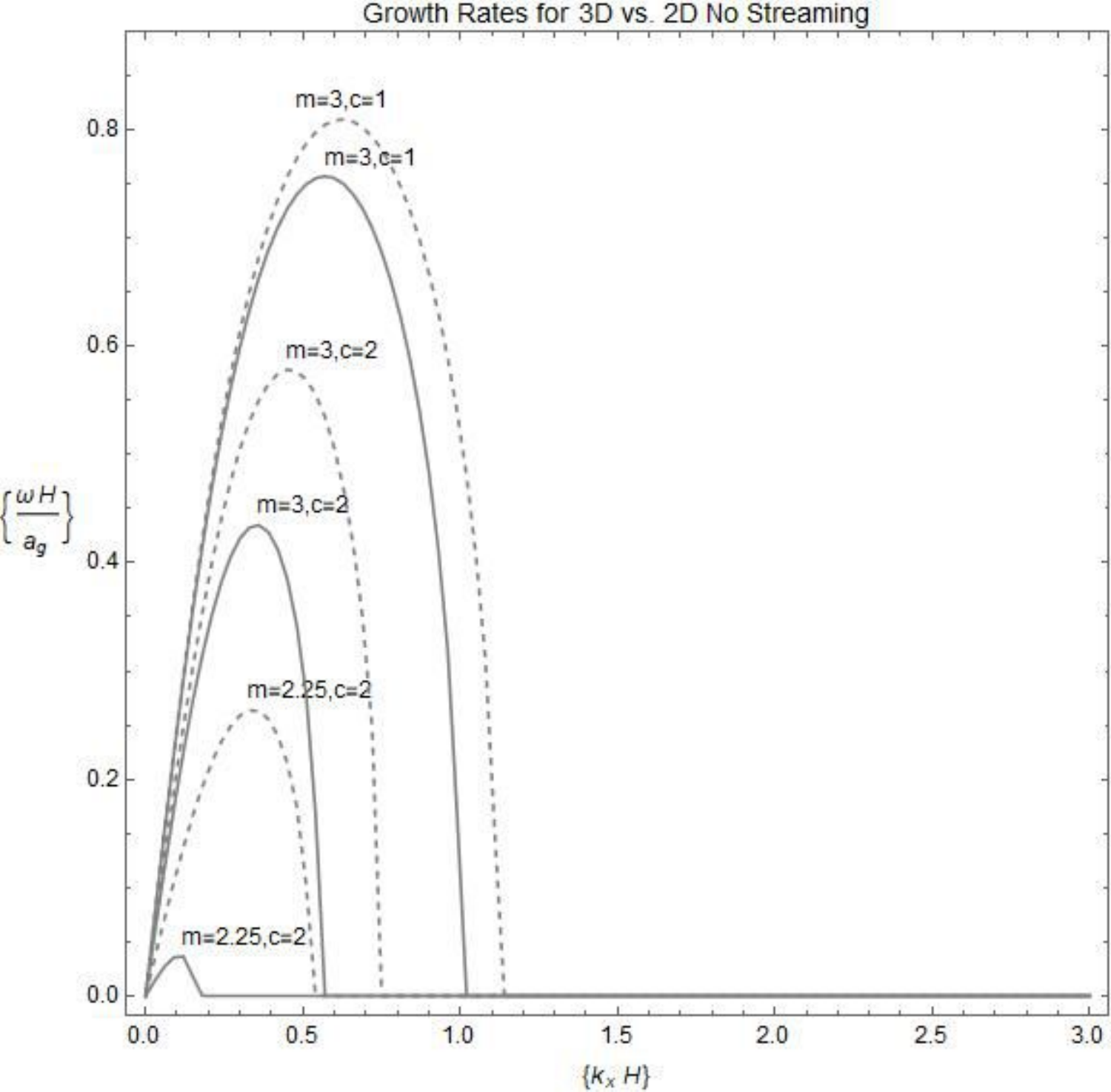}{0.37\textwidth}{(b)}
		 }
\caption{These plots shows the difference in growth rate for the 2D and 3D cases of the Modified Parker with Streaming case and the Modified Parker case Note we have chosen three different combinations of $m$ and $c$ values. The dashed line is the 3D case, while the solid line is the 2D case. For the 2D case, $k_y=0$ and for the 3D case, $k_y\rightarrow\infty$. Note that the 2D case is always unstable over a smaller domain than the 3D case.
\label{fig:2d3d}}
\end{figure}

In Figure \ref{fig:2d3d}, we have plotted the 2D maximum growth rate versus the 3D maximum growth rate for the Modified Parker with Streaming case and Modified Parker case in terms of the same dimensionless units. The results of these plots match the conclusions of \cite{newcombinstability1961} where the most unstable mode had $k_y\rightarrow\infty$ with a total pressure perturbation of zero. For all three values of $m$ and $c$ in the plots, we see that the 3D case is always unstable over a larger range of wavenumbers. However, while for the Modified Parker case the 2D case always has a lower maximum growth rate than the 3D case, for the Modified Parker with Streaming case the maximum growth rate is actually larger in the 2D case. This is a result of the reduced role of pressure in driving the 3D instability.

\section{Summary \& Conclusions}\label{sec:summary}
Cosmic rays have recently emerged as one of the ``big three" agents of star formation feedback, the other two being thermal pressure and radiation pressure. Unlike thermal gas, cosmic rays cool relatively slowly, and up to now there has been no evidence that cosmic rays, unlike radiation, preferentially escape along  the lowest density paths.

In this paper, we have aimed to determine the effects of
the cosmic ray equation of state and cosmic ray streaming on the Parker Instability. In essence, this is an
exploratory study to assess the role of cosmic rays in sculpting their own environment. It should be viewed as complementary to studies such as \cite{Farber2017} and \cite{Wiener2017}, which explore the coupling between cosmic rays and interstellar gas in different phases. 

Here, we give a review of our main results.

The original result of Parker in 2D treated the cosmic rays as a $\gamma_c=0$ fluid. We found that by modifying the treatment to let $\gamma_c=4/3$, the system becomes much more stable, as argued qualitatively in \cite{zweibelparker1975} and \cite{boettcherEDIG2016}. This is due to the extra work required to compress the cosmic rays as they flow into the magnetic valleys. The result is that the puffed up galactic disks seen in models without cosmic ray transport (\cite{uhligwinds2012,ruszkowskiwinds2017}) may be more stable than the classic treatment of the Parker Instability would imply.

However, once cosmic ray streaming at the Alfv\'{e}n speed was introduced, the system becomes extremely unstable over a much larger range of wavenumbers. With cosmic ray streaming, the instability can grow about $100\times$ faster at a wavelength that is about $20\times$ smaller than it does in the modified system as long as the thermal gas has an adiabatic index  $\gamma_g = 5/3$. If
$\gamma_g = 1$, streaming has a much smaller effect (Figure \ref{fig:parkerStream}). We attribute the difference to the destabilizing effect of
cosmic ray heating. This is a much shorter wavelength instability than Classic Parker, and requires numerical resolution below 100pc to appear. This result also shows the importance of a full thermodynamic treatment of the gas that includes heating and cooling beyond simply prescribing a polytropic index.

Comparison of the three Parker Instability cases, along with analysis of the work contributions compared with the overstability of acoustic waves, further indicates that cosmic ray heating is the destabilizing effect.

Similar analysis of the same cases in 3D for $k_y\rightarrow\infty$ allowed us to compare its results to our 2D cases. The 3D case was found to be unstable over a larger range of wavelengths than the 2D case. These two cases together give us our upper and lower limits on the range of unstable modes and the maximum growth rate for the Parker Instability with Streaming.

Based on these results, we believe that the Parker Instability, with cosmic ray streaming,  may have a non-negligible effect in star forming galaxies. Therefore, it may be important to carry out simulations at sufficient resolution to see if the puffed-up discs found in some of these simulations are truly stable. 
 
However, this analysis of the instability is not fully complete. A full analysis would include diffusion,  thermal conduction, and radiative cooling. These topics, along with nonlinear simulations, will be included in future work. 

\acknowledgments
We would like to thank the referee for their very insightful comments. We gratefully acknowledge useful discussions with Chad Bustard, Jacqueline Goldstein, and Josh Wiener as well as the support of the Vilas Trust and WARF Foundation at the University of Wisconsin and NSF Grant AST 1616037. EGZ appreciates the hospitality of the University of Chicago, where a portion of this work was carried out.

\appendix 
\section{Modified Parker Dispersion Relations}\label{sec:modparkdisp}
We now provide the dispersion relations for both the Modified Parker and the Modified Parker with Streaming cases. First, in order to allow for easy comparison, we rewrite the dispersion relations for the overstability of acoustic waves and for Classic Parker from eqns. \ref{eq:acousticdisp} and \ref{eq:classicparkdisp}. 
The acoustic dispersion relation is:
\begin{equation}
\omega(\omega-k u_A)(\omega^2-k^2 a_g^2)
-k^2 a_c^2(\omega - k u_A/2)(\omega + (\gamma_g-1)k u_A) = 0
\end{equation}
The Classic Parker dispersion relation is:
\begin{equation}
\gamma_g^2\hat{\omega}^4-\hat{\omega}^2(\gamma_g^2+2\alpha\gamma_g)(\hat{k_x}^2+\hat{k_z}^2+\frac{1}{4})+\hat{k_x}^2\Big(2\alpha\gamma_g(\hat{k_x}^2+\hat{k_z}^2)
-2(\alpha+\beta+\frac{1}{2})-(\alpha+\beta)^2+\gamma_g(\frac{3\alpha}{2}+\beta+1)\Big) = 0
\end{equation}
using the same dimensionless definitions as used in eqns. \ref{eq:mc}, \ref{eq:q}, \ref{eq:dimendefn}, and \ref{eq:newpertpress}.
For Modified Parker, using the dimensionless eqns. \ref{eq:modparkcont}-\ref{eq:modparkgas}, we are able to gain the dispersion relation:
\begin{equation}\label{eq:modparkerdisp}
\begin{split}
i\hat{\omega}^7 \Big(-4 c^2 \gamma_c \gamma_g \hat{k_x}^2 q-4 \gamma_c \gamma_g \hat{k_z}^2 q \Big(c^2+m^2+1\Big)-2 c^2 \gamma_g (-1+2 i \hat{k_z})q-c^2 \gamma_c \gamma_g q \\
-4 \gamma_c \gamma_g \hat{k_x}^2 m^2 q-4 \gamma_c \gamma_g \hat{k_x}^2 q
+2 \gamma_c \gamma_g (-1+2 i\hat{k_z})-2 i \gamma_c \hat{k_z} q \Big(\gamma_g m^2+2\Big)-\gamma_c \gamma_g q+2 \gamma_c q\Big) \\
+i\hat{\omega}^5 \Big(4 \Big(c^2+1\Big) \gamma_c \gamma_g \hat{k_x}^4 m^2 q+4 c^2 \gamma_c \gamma_g \hat{k_x}^2-4 c^2 \gamma_g \hat{k_x}^2+c^2 \gamma_c \gamma_g \hat{k_x}^2 m^2 \Big(4 \hat{k_z}^2 q+q\Big)+4 \gamma_c \gamma_g \hat{k_x}^2 \\
-4 \gamma_c \hat{k_x}^2+4 \gamma_c \gamma_g \hat{k_x}^2 \hat{k_z}^2 m^2 q-2 \gamma_c \gamma_g \hat{k_x}^2 m^2+\gamma_c \gamma_g \hat{k_x}^2 m^2 q\Big) \\
+4i \gamma_c \gamma_g q \hat{\omega}^3 = 0
\end{split}
\end{equation}
where each term is defined as before according to eqns. \ref{eq:mc}, \ref{eq:q}, \ref{eq:dimendefn}, and \ref{eq:newpertpress}. We have organized the dispersion relation based on the power of $\hat{\omega}$ associated with each term. Note that there are three roots for which there is no propagation and are trivial.

We can also get the dispersion relation for the Modified Parker with Streaming case. We now use the dimensionless eqns. \ref{eq:parkstreamcont} and \ref{eq:parkstreamgas}. The resulting dispersion relation is:
\begin{equation}\label{eq:parkerstreamdisp}
\begin{split}
8 \gamma_c \gamma_g q (i\hat{\omega}^7 - i \hat{k_x} m\hat{\omega}^6) \\
+i\hat{\omega}^5 \Big(-8 c^2 \gamma_c \gamma_g \hat{k_x}^2 q-2 c^2 \gamma_g q \Big(\gamma_c+4 \gamma_c \hat{k_z}^2 + 4 i \hat{k_z}-2\Big)-4 \gamma_c \gamma_g-8 \gamma_c \gamma_g \hat{k_x}^2 m^2 q-8 \gamma_c \gamma_g \hat{k_x}^2 q \\
-8 \gamma_c \gamma_g \hat{k_z}^2 \Big(m^2+1\Big) q+8 i \gamma_c \gamma_g \hat{k_z}-4 i \gamma_c \hat{k_z} q \Big(\gamma_g m^2+2\Big)-2 \gamma_c
\gamma_g q+4 \gamma_c q\Big) \\
+i\hat{\omega}^4 \Big(-4 c^2 \gamma_c \gamma_g (2 \gamma_g-3) \hat{k_x}^3 m q+c^2 \gamma_g \hat{k_x} m q \Big(\gamma_c (3-2 \gamma_g)+4\gamma_c (3-2 \gamma_g) \hat{k_z}^2+8 i \hat{k_z}-4\Big)+8 \gamma_c \gamma_g \hat{k_x}^3 m^3 q 
+8 \gamma_c \gamma_g \hat{k_x}^3 m q \\ 
+ 8\gamma_c \gamma_g \hat{k_x} \hat{k_z}^2 m \Big(m^2+1\Big) q+4 i \gamma_c \hat{k_x} \hat{k_z} m q \Big(\gamma_g m^2+2\Big)
-8 i \gamma_c\gamma_g \hat{k_x} \hat{k_z} m+4 \gamma_c \gamma_g \hat{k_x} m+2 \gamma_c \gamma_g \hat{k_x} m q-4 \gamma_c \hat{k_x} m q\Big) \\
+i\hat{\omega}^3 \Big(4 c^2 \gamma_c \gamma_g^2 \hat{k_x}^4 m^2 q+4 c^2 \gamma_c \gamma_g \hat{k_x}^4 m^2 q+8 c^2 \gamma_c \gamma_g \hat{k_x}^2-8 c^2 \gamma_g \hat{k_x}^2+c^2 \gamma_c \gamma_g (\gamma_g+1) \hat{k_x}^2 \Big(4 \hat{k_z}^2+1\Big) m^2 q \\
+8 \gamma_c\gamma_g \hat{k_x}^4 m^2 q+8 \gamma_c \gamma_g \hat{k_x}^2-8 \gamma_c \hat{k_x}^2+8 \gamma_c \gamma_g \hat{k_x}^2 \hat{k_z}^2 m^2 q-4 \gamma_c \gamma_g \hat{k_x}^2 m^2+2 \gamma_c \gamma_g \hat{k_x}^2 m^2 q\Big) \\
+i\hat{\omega}^2 \Big(4 c^2 \gamma_c \gamma_g (2 \gamma_g-3) \hat{k_x}^5 m^3 q+c^2 \gamma_c \gamma_g (2 \gamma_g-3) \hat{k_x}^3 m^3 \Big(4 \hat{k_z}^2 q+q\Big)+4 c^2 \gamma_c \gamma_g (2 \gamma_g-3) \hat{k_x}^3 m+8 c^2 \gamma_g \hat{k_x}^3 m \\
-8 \gamma_c \gamma_g \hat{k_x}^5 m^3 q-8 \gamma_c \gamma_g \hat{k_x}^3 \hat{k_z}^2 m^3 q+4 \gamma_c \gamma_g \hat{k_x}^3 m^3-2 \gamma_c \gamma_g \hat{k_x}^3 m^3 q-8 \gamma_c \gamma_g \hat{k_x}^3 m+8 \gamma_c \hat{k_x}^3 m\Big) \\
+i\hat{\omega} \Big(-4 c^2 \gamma_c (\gamma_g-1) \gamma_g \hat{k_x}^6 m^4 q-c^2 \gamma_c (\gamma_g-1) \gamma_g \hat{k_x}^4 \Big(4 \hat{k_z}^2+1\Big) m^4 q-4 c^2 \gamma_c \gamma_g^2 \hat{k_x}^4 m^2+4 c^2 \gamma_c
\gamma_g \hat{k_x}^4 m^2\Big) = 0
\end{split}
\end{equation}
where again each term is defined according to eqns. \ref{eq:mc}, \ref{eq:q}, \ref{eq:dimendefn}, and \ref{eq:newpertpress}. We have also organized this dispersion relation by the power of $\hat{\omega}$ associated with each term. Again, note for this case that there is one mode which has no propagation and is trivial.

Finally, we can also get the dispersion relation for Modified Parker with Cosmic Ray Streaming in a 3D analysis where $k_y\rightarrow \infty$ and the total pressure perturbation goes to zero. This dispersion relation is:
\begin{equation}
\begin{split}
\hat{\omega}(\hat{\omega-m\hat{k_x}})[(\hat{\omega}^2-m^2\hat{k_x}^2-\frac{1}{q})\{(\hat{\omega}^2-m^2\hat{k_x}^2)(1+m^2)+m^4\hat{k_x}^2\}+\frac{1}{q^2}(\hat{\omega}^2-m^2\hat{k_x}^2)]\\
+c^2(\hat{\omega}^2-m^2\hat{k_x}^2)(\hat{\omega}^2-m^2\hat{k_x}^2-\frac{1}{q})(\hat{\omega}-\frac{m\hat{k_x}}{2})(\hat{\omega}+(\gamma_g-1)m\hat{k_x}) = 0
\end{split}
\end{equation}
where the dimensionless quantities are defined the same as in the 2D analysis.
\\ \\ \\ \\ \\ \\ \\ \\ \\ \\ \\ \\ \\ \\ \\ \\ \\ \\ \\ 


\pagebreak
\bibliographystyle{aasjournal}
\bibliography{citations}

\begin{thebibliography}{}
\expandafter\ifx\csname natexlab\endcsname\relax\def\natexlab#1{#1}\fi
\providecommand{\url}[1]{\href{#1}{#1}}

\bibitem[{{Begelman} \& {Zweibel}(1994)}]{begelmanacoustic1994}
{Begelman}, M.~C., \& {Zweibel}, E.~G. 1994, \apj, 431, 689

\bibitem[{{Boettcher} {et~al.}(2016){Boettcher}, {Zweibel}, {Gallagher}, \&
  {Benjamin}}]{boettcherEDIG2016}
{Boettcher}, E., {Zweibel}, E.~G., {Gallagher}, III, J.~S., \& {Benjamin},
  R.~A. 2016, \apj, 832, 118

\bibitem[{{Booth} {et~al.}(2013){Booth}, {Agertz}, {Kravtsov}, \&
  {Gnedin}}]{boothwinds2013}
{Booth}, C.~M., {Agertz}, O., {Kravtsov}, A.~V., \& {Gnedin}, N.~Y. 2013,
  \apjl, 777, L16

\bibitem[{{Breitschwerdt} {et~al.}(1991){Breitschwerdt}, {McKenzie}, \&
  {Voelk}}]{breitschwerdtwinds1991}
{Breitschwerdt}, D., {McKenzie}, J.~F., \& {Voelk}, H.~J. 1991, \aap, 245, 79

\bibitem[{{Drury} \& {Falle}(1986)}]{druryacoustic1986}
{Drury}, L.~O., \& {Falle}, S.~A.~E.~G. 1986, \mnras, 223, 353

\bibitem[{{Farber} {et~al.}(2017){Farber}, {Ruszkowski}, {Yang}, \&
  {Zweibel}}]{Farber2017}
{Farber}, R., {Ruszkowski}, M., {Yang}, H.-Y.~K., \& {Zweibel}, E.~G. 2017,
  ArXiv e-prints, arXiv:1707.04579

\bibitem[{{Farmer} \& {Goldreich}(2004)}]{Farmer2004}
{Farmer}, A.~J., \& {Goldreich}, P. 2004, \apj, 604, 671

\bibitem[{{Grenier} {et~al.}(2015){Grenier}, {Black}, \&
  {Strong}}]{Grenier2015}
{Grenier}, I.~A., {Black}, J.~H., \& {Strong}, A.~W. 2015, \araa, 53, 199

\bibitem[{{Guo} {et~al.}(2012){Guo}, {Mathews}, {Dobler}, \&
  {Oh}}]{guofermi2012}
{Guo}, F., {Mathews}, W.~G., {Dobler}, G., \& {Oh}, S.~P. 2012, \apj, 756, 182

\bibitem[{{Hanasz} {et~al.}(2009){Hanasz}, {Otmianowska-Mazur}, {Kowal}, \&
  {Lesch}}]{Hanasz2009}
{Hanasz}, M., {Otmianowska-Mazur}, K., {Kowal}, G., \& {Lesch}, H. 2009, \aap,
  498, 335

\bibitem[{{Kulsrud} \& {Pearce}(1969)}]{Kulsrud1969}
{Kulsrud}, R., \& {Pearce}, W.~P. 1969, \apj, 156, 445

\bibitem[{{Kulsrud}(2005)}]{KulsrudBook2005}
{Kulsrud}, R.~M. 2005, {Plasma physics for astrophysics}

\bibitem[{{Kuwabara} \& {Ko}(2006)}]{kuwabaraparker2006}
{Kuwabara}, T., \& {Ko}, C.-M. 2006, \apj, 636, 290

\bibitem[{{Kuwabara} {et~al.}(2004){Kuwabara}, {Nakamura}, \&
  {Ko}}]{kuwabaraparker2004}
{Kuwabara}, T., {Nakamura}, K., \& {Ko}, C.~M. 2004, \apj, 607, 828

\bibitem[{{Newcomb}(1961)}]{newcombinstability1961}
{Newcomb}, W.~A. 1961, Physics of Fluids, 4, 391

\bibitem[{{Parker}(1966)}]{parkerinstability1966}
{Parker}, E.~N. 1966, \apj, 145, 811

\bibitem[{{Ruszkowski} {et~al.}(2017){Ruszkowski}, {Yang}, \&
  {Zweibel}}]{ruszkowskiwinds2017}
{Ruszkowski}, M., {Yang}, H.-Y.~K., \& {Zweibel}, E. 2017, \apj, 834, 208

\bibitem[{{Ryu} {et~al.}(2003){Ryu}, {Kim}, {Hong}, \& {Jones}}]{ryuparker2003}
{Ryu}, D., {Kim}, J., {Hong}, S.~S., \& {Jones}, T.~W. 2003, \apj, 589, 338

\bibitem[{{Salem} \& {Bryan}(2014)}]{Salem2014}
{Salem}, M., \& {Bryan}, G.~L. 2014, \mnras, 437, 3312

\bibitem[{{Santill{\'a}n} {et~al.}(2000){Santill{\'a}n}, {Kim}, {Franco},
  {Martos}, {Hong}, \& {Ryu}}]{Santillan2000}
{Santill{\'a}n}, A., {Kim}, J., {Franco}, J., {et~al.} 2000, \apj, 545, 353

\bibitem[{{Uhlig} {et~al.}(2012){Uhlig}, {Pfrommer}, {Sharma}, {Nath},
  {En{\ss}lin}, \& {Springel}}]{uhligwinds2012}
{Uhlig}, M., {Pfrommer}, C., {Sharma}, M., {et~al.} 2012, \mnras, 423, 2374

\bibitem[{{Wiener} {et~al.}(2017){Wiener}, {Oh}, \& {Zweibel}}]{Wiener2017}
{Wiener}, J., {Oh}, S.~P., \& {Zweibel}, E.~G. 2017, \mnras, 467, 646

\bibitem[{{Yang} {et~al.}(2012){Yang}, {Ruszkowski}, {Ricker}, {Zweibel}, \&
  {Lee}}]{yangfermi2012}
{Yang}, H.-Y.~K., {Ruszkowski}, M., {Ricker}, P.~M., {Zweibel}, E., \& {Lee},
  D. 2012, \apj, 761, 185

\bibitem[{{Zweibel}(2017)}]{zweibelreview2017}
{Zweibel}, E.~G. 2017, Physics of Plasmas, 24, 055402

\bibitem[{{Zweibel} \& {Kulsrud}(1975)}]{zweibelparker1975}
{Zweibel}, E.~G., \& {Kulsrud}, R.~M. 1975, \apj, 201, 63

\end{thebibliography}





\end{document}